\documentclass{article}

\PassOptionsToPackage{numbers,sort&compress}{natbib}
\usepackage[preprint]{neurips_2026}

\usepackage[utf8]{inputenc}
\usepackage[T1]{fontenc}
\usepackage{hyperref}
\usepackage{url}
\usepackage{booktabs}
\usepackage{amsfonts}
\usepackage{nicefrac}
\usepackage{microtype}
\usepackage{xcolor}
\usepackage{amsmath,amssymb}
\usepackage{graphicx}
\usepackage{algorithm}
\usepackage{algpseudocode}
\usepackage[capitalize,noabbrev]{cleveref}

\newcommand{\sys}{HiSparse}
\newcommand{\kv}{KV}
\newcommand{\hbm}{HBM}
\newcommand{\cpu}{CPU}
\newcommand{\gpu}{GPU}

\newcommand{\paragraphhead}[1]{\vspace{2pt}\noindent\textbf{#1.}}

\title{\sys: Scaling Sparse-Attention Decoding with Hierarchical KV Cache Management}

\author{%
  Zhiqiang Xie\thanks{Corresponding author: \texttt{xiezhq@cs.stanford.edu}} \\
  Stanford University \& Meta \\
  \And
  Zhangheng Huang \\
  Alibaba Group\\
  \And
  Tingwei Huang \\
  Ant Group\\
  \AND
  Ziyi Xu \\
  Shanghai Jiao Tong University \\
  \And
  Ruiyang Ma \\
  Peking University \\
  \And
  Christos Kozyrakis \\
  Stanford University \& NVIDIA Research \\
}

\begin{document}
\maketitle

\begin{abstract}
Top-$k$ sparse attention makes long-context LLM decoding cheap to
compute: each step reads only a few thousand selected KV entries
rather than the full context. Serving systems, however, typically keep
the entire KV cache in GPU HBM so that every position stays
selectable, so a request's memory bill still grows with its full
context length---decoding hits a capacity wall long before it runs out
of compute, and a context whose KV cache exceeds HBM cannot be served
at all. We present \sys{}, an exact, indexer-agnostic hierarchical KV
cache for sparse-attention serving. \sys{} keeps each request's full
KV history in host memory and bounds its decode footprint with a
small, fixed-size GPU cache; a fused CUDA kernel resolves each layer's
selections---hit detection, LRU replacement, and host-to-device
fetches---inside the decode CUDA graph; and, for models that share
selections across layers, exact layer-wise prefetching hides roughly
half of the remaining miss overhead. Because only KV placement
changes, model outputs are unchanged. \sys{} is merged into upstream
SGLang and evaluated across three sparse-attention families (DSA, NSA,
and Quest) on H200, B200, and GH200 platforms: it improves peak
generation throughput by up to $4.7\times$ on long-context workloads
while preserving comparable per-token latency and reducing
time-to-first-token at high load---and a no-IO oracle shows the
resolution mechanism itself adds no measurable per-token cost, leaving
host--device IO as the only price of bounded residency.
\end{abstract}

\section{Introduction}
\label{sec:intro}

Long-context inference is becoming a standard LLM workload: coding
agents inspect entire repositories, assistants synthesize across long
documents, and recent models target context windows from tens of
thousands to millions of
tokens~\citep{deepseek_v4_2026,glm5_2026,qwen3_30b_a3b_thinking_2507}.
Serving these contexts remains expensive, because each live decoding
request carries a KV cache that grows with its full history.

Top-$k$ sparse attention offers a promising path to scaling long
contexts. Instead of attending to every previous token, each decode
step attends to a small, query-dependent set of $k$ selected KV
entries---typically a few thousand tokens, one to two orders of
magnitude below the contexts these models target. DeepSeek-V3.2
demonstrates that a learned top-$k$ selection preserves model quality
while sharply reducing long-context attention
cost~\citep{dsa2025,deepseek_v32_hf_2025}; the same pattern appears in
trained architectures such as DeepSeek Sparse Attention (DSA) and
Native Sparse Attention (NSA)~\citep{yuan2025nsa} and in training-free
selectors such as Quest~\citep{tang2024quest}. Once the selected set
is fixed, the attention kernel reads only $k$ KV entries rather than
the full context, suggesting that long-context decoding should become
much cheaper to serve.

\begin{figure}[t]
\centering
\includegraphics[width=0.95\linewidth]{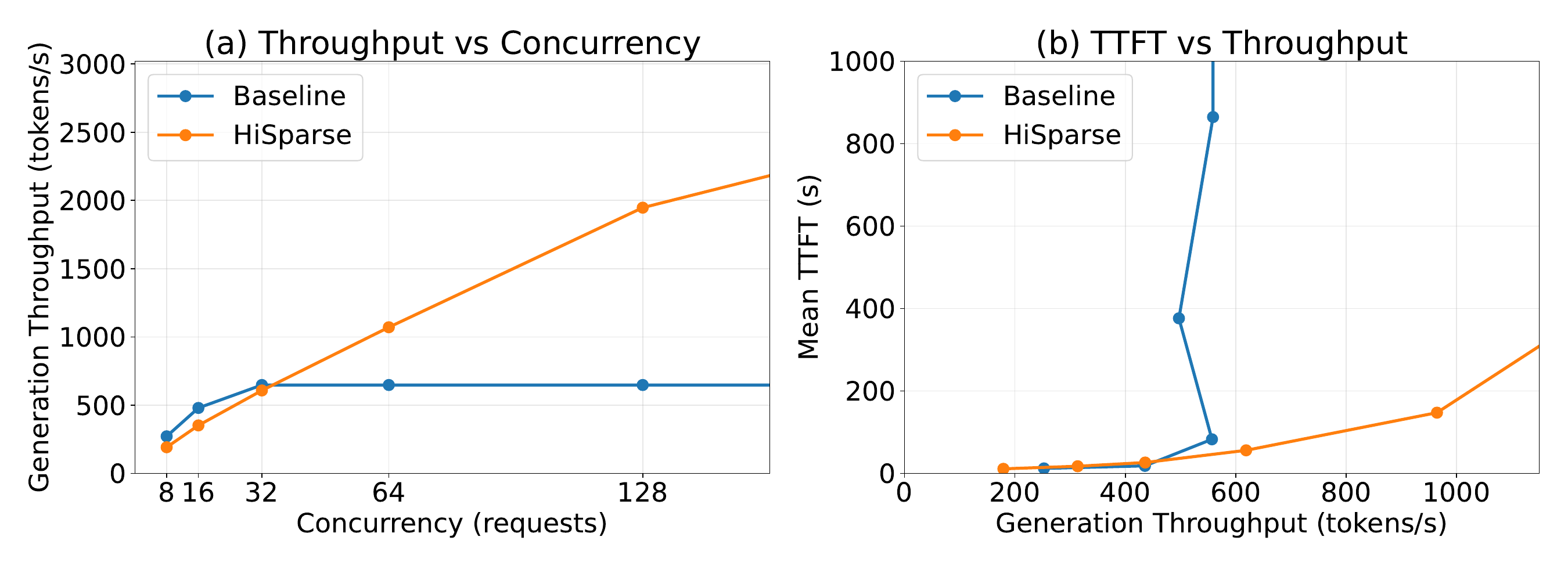}
\caption{\sys{} decouples decoding throughput from GPU memory
capacity on long-context workloads.
\textbf{(a)}~Decoding throughput vs.\ concurrency: the baseline
plateaus once HBM saturates; \sys{} continues to scale. Longer
contexts hit the same wall at proportionally lower concurrency.
\textbf{(b)}~Mean TTFT vs.\ throughput in PD-colocated mode:
\sys{} sustains lower TTFT at higher throughput.
GLM-5.1-FP8 (DSA, $k{=}2048$), $8{\times}$H200, $32$K input, $8$K
output.}
\label{fig:motivation}
\end{figure}

Sparse attention, however, does not shrink the KV capacity
bottleneck. The selected set changes across generated tokens and
layers---an entry skipped now may be selected later---so serving
systems typically keep the full KV cache resident in GPU HBM to keep
every logical position addressable. The result is a lopsided serving
economics: each decode step reads only $k$ entries, yet every entry
of the full context pays for HBM residency so that it \emph{might} be
read---attention became orders of magnitude cheaper, while its memory
bill did not drop by a byte. The bill is substantial even for
compressed KV layouts: a $128$K-token GLM-5.1 request holds
$13.09$\,GB of BF16 KV state, and a single $1$M-token request would
claim an entire H200---more than $100$\,GB of its $141$\,GB HBM
before counting weights; with weights resident, it cannot be admitted
at all, capping the servable context well below the model's window.
In practice the wall binds on the
\emph{product} of context length and concurrency: at $32$K tokens per
request, a few dozen concurrent requests exhaust HBM
(Figure~\ref{fig:motivation}), and at $128$K the same HBM admits
$4\times$ fewer. Long-context sparse decoding therefore runs out of
memory capacity long before it runs out of attention compute.

Figure~\ref{fig:motivation} shows the serving consequence: decoding
throughput plateaus once HBM fills, and in colocated serving the
time-to-first-token (TTFT) climbs as decode KV caches crowd out
prefill work. Yet full-KV serving ignores a free oracle. The sparse
selector announces, at every layer of every decode step, exactly
which $k$ positions attention will read---a precise, per-step demand
signal for KV placement that no dense-attention system ever had. That demand also has structure: consecutive decode steps
reselect largely overlapping positions~\citep{chen2025ess}, and nearby
layers select correlated positions, a locality that recent models make
explicit by sharing indexer outputs across
layers~\citep{bai2026indexcache,glm52_blog_2026}. Sparse selections
behave like memory accesses with strong temporal locality---exactly
the regime in which a small fast cache backed by a larger, slower tier
works. This observation motivates \sys{}, an exact hierarchical KV
cache system for top-$k$ sparse-attention serving: HBM should pay for
what attention reads, not for everything it might read.

\sys{} keeps each request's complete KV history in host memory and
gives it a small, fixed-size \emph{GPU cache} in HBM. Each layer's
top-$k$ selections are resolved against this cache before its
attention kernel launches: resident entries are used in place, and the
few that miss are fetched from the host copy in one batched
transfer. Selected positions, attention
scores, and outputs are unchanged, and because \sys{} consumes only
the selected positions each layer emits, it is
\emph{indexer-agnostic}: it slots beneath DSA, NSA, or Quest without
retraining or model changes. A request's decode-time HBM consumption
therefore scales with the GPU-cache size rather than with its context
length.

The crux of \sys{} is keeping this hierarchy off the decode critical
path; its design contributions answer three challenges. First,
\emph{what to keep resident}: staging only the current top-$k$ set
discards locality---on a LongBenchV2 selection trace it misses $30\%$
of each step's selections---so \sys{} manages the cache with LRU,
turning selection locality into an $87\%$ hit rate at twice the
top-$k$ size (\S\ref{sec:eval-cache}). Second, \emph{hiding the
misses that remain}: each miss stalls that layer's attention on a
host-memory fetch, and at high batch size the misses of concurrent
requests contend for the same host link (\S\ref{sec:eval-micro});
\sys{} makes each fetch bandwidth-efficient with GPU-assisted IO and,
when models share selections across layers, overlaps the remaining
transfers with the intervening layers' computation via exact prefetch
(\S\ref{sec:prefetch}). Third, \emph{making resolution itself cheap}:
hit detection, victim selection, metadata updates, and host fetches
recur at every sparse layer of every decode step, so \sys{} fuses them
into a single CUDA kernel captured inside the decode CUDA graph
(\S\ref{sec:kernel}).

\sys{} is not a research prototype: our implementation is merged into
upstream SGLang~\citep{zheng2024sglang}, a widely deployed open-source
serving framework, and released as a supported serving
feature~\citep{hisparse_blog_2026,sglang_hisparse_docs_2026}
(Appendix~\ref{sec:appendix-impl} details the integration). We
evaluate it across three sparse-attention families (DSA, NSA, and
Quest) on three hardware platforms (H200, B200, and GH200)\@. \sys{}
improves peak generation throughput by up to $4.7\times$ on
long-context workloads, preserves a comparable time per output token
(TPOT) in the overlapping throughput range, reduces TTFT at high
load, and does so without changing model outputs. All of these gains
come from one lever: bounded residency lets \sys{} run much larger
decode batches in the same HBM. The same lever can be pulled the
other way---serving a fixed batch with substantially less HBM
(\S\ref{sec:eval-end2end}), or serving contexts that HBM alone could
never hold (\S\ref{sec:limitations}).

This paper makes the following contributions:
\begin{itemize}
\item We identify the capacity wall in long-context top-$k$
sparse-attention serving: active KV reads scale with $k$, but the
HBM-resident KV footprint still scales with the full context length
(\S\ref{sec:background}).
\item We introduce \sys{}, an exact, indexer-agnostic hierarchical KV
cache that keeps full KV state available in host memory while bounding
per-request decode HBM with a fixed-size GPU cache
(\S\ref{sec:design}).
\item We design a locality-preserving miss-resolution path: LRU
management that turns selection locality into cache hits, layer-wise
prefetching that hides residual miss latency, and a fused CUDA resolve
kernel that keeps resolution cheap on the decode critical path
(\S\ref{sec:kernel}, \S\ref{sec:prefetch}).
\item We evaluate \sys{} on DSA, NSA, and Quest workloads across H200,
B200, and GH200 platforms, showing large long-context throughput gains
and analyzing the cache-policy, kernel, and host-device bandwidth
tradeoffs that determine performance (\S\ref{sec:eval}).
\end{itemize}

\section{Background and Motivation}
\label{sec:background}
\subsection{Top-$k$ Sparse Attention}
\label{sec:sparse-attn}

Top-$k$ sparse attention replaces full attention over the entire
context with attention over a small, query-dependent set of keys. At
decode step $t$, an \emph{indexer} produces a selected set
$\mathcal{S}_t \subseteq \{1,\dots,L_{\text{ctx}}\}$ with
$|\mathcal{S}_t| = k$, and the attention kernel reads only the
corresponding key and value entries. The resulting sparse attention is
still exact with respect to the model's sparse-attention rule: once
$\mathcal{S}_t$ is fixed, unselected entries do not participate in the
current attention computation.

Recent systems differ mainly in how they produce $\mathcal{S}_t$
(Table~\ref{tab:selectors}). DeepSeek Sparse Attention (DSA)
introduces a learned ``lightning indexer'' co-trained with the
backbone and used by DeepSeek-V3.2 and
GLM-5.1~\citep{dsa2025,deepseek_v32_hf_2025,glm5_2026,hisparse_blog_2026}: it keeps one compact indexer key
per token and scores the full history with a lightweight kernel before
taking a token-level top-$k$. Native Sparse Attention (NSA) combines
compressed, selected, and sliding-window branches in a trainable
architecture; its selected branch keeps one compressed key per block
of tokens and takes a top-$k$ over block scores~\citep{yuan2025nsa}.
Quest is a training-free, page-granular selector for dense pretrained
models: it maintains per-page min/max summaries of keys and selects
the pages with the highest query-aware upper
bounds~\citep{tang2024quest}. The selectors thus differ in the state
they maintain, the scoring they perform, and their selection
granularity---token, block, or page.

\begin{table}[t]
\centering
\caption{Query-dependent top-$k$ selectors. Each maintains compact,
HBM-resident selection state and emits logical token positions; the
KV records read by attention dominate memory and are what \sys{}
manages hierarchically.}
\label{tab:selectors}
\small
\begin{tabular}{@{}llll@{}}
\toprule
Selector & Selection state (memory) & Selection compute & Notes \\
\midrule
DSA~\citep{dsa2025} & compact indexer key per token & score history, top-$k$ tokens & token-level; co-trained \\
NSA~\citep{yuan2025nsa} & compressed key per token block & score blocks, top blocks & block-level; trained \\
Quest~\citep{tang2024quest} & min/max key vectors per page & bound pages, top pages & page-level; training-free \\
\bottomrule
\end{tabular}
\end{table}

Despite these differences, the selectors share three properties that
define a common systems interface. First, the state needed to
\emph{choose} is compact: indexer keys, block keys, and page summaries
are all far smaller than the KV records that attention reads, so
selection state can stay HBM-resident even at long context. Second,
selection completes \emph{before} the attention kernel touches any KV
record, creating a natural interposition point between choosing and
reading. Third, the output has the same form everywhere: a set of
logical token positions per layer, which we write
$\mathcal{S}_t^{(\ell)}$ for layer $\ell$ at step $t$. Each sparse
layer ordinarily selects its own set, though recent models amortize
indexer cost by letting groups of consecutive layers \emph{share} one
layer's selection~\citep{bai2026indexcache,glm52_blog_2026}---a design
\sys{} later exploits for prefetching (\S\ref{sec:prefetch}). \sys{}
builds on exactly these properties (\S\ref{sec:design}): it leaves selection state and
computation untouched on the GPU, interposes at the position
interface, and applies hierarchical placement only to the bulky KV
records---which is what makes it indexer-agnostic.

\subsection{KV Cache and the Capacity Wall}
\label{sec:capacity-wall}

During autoregressive decoding, the keys and values of every previous
token are retained as the KV cache. Top-$k$ sparse attention narrows
what one step \emph{reads}, not what must remain \emph{available}:
$\mathcal{S}_t$ is drawn from the full history and drifts from step to
step, so an entry skipped now may be selected later, and serving
systems typically keep the entire cache HBM-resident to keep every
position addressable by the indexer and attention backend
(\S\ref{sec:intro}). The resulting admission constraint is the
quantitative form of the capacity wall: a decode batch of
$N_{\text{batch}}$ requests at context length $L_{\text{ctx}}$ must
fit $N_{\text{batch}} \times L_{\text{ctx}}$ tokens of KV state in
whatever HBM remains after model weights, while each decode step's
attention reads only $N_{\text{batch}} \times k$ of those tokens.

The wall hits both common deployment modes. Adding concurrent requests
improves throughput only until KV storage fills HBM, after which the
scheduler cannot admit more decode work even though the attention
kernels have compute headroom---the plateau in
Figure~\ref{fig:motivation}(a). In \emph{PD-disaggregated} serving, where prefill
and decode run on separate GPU pools, HBM capacity directly caps
decode-pool throughput. In \emph{PD-colocated} serving, prefill and
decode share GPUs and HBM: full-context decode KV caches consume
memory that prefill chunks need, so incoming requests wait for both
memory and decode slots before producing their first token, and mean
TTFT becomes dominated by queueing and rises sharply with load
(Figure~\ref{fig:motivation}(b)).

Our GLM-5.1 deployment (\S\ref{sec:eval-setup}: $8\times$H200,
$1.1$\,TB aggregate HBM) makes the numbers concrete. At $32$K input /
$8$K output, each request holds up to $4$\,GB of KV state, and the
full-KV baseline saturates at about $60$ concurrent requests---roughly
$240$\,GB of KV, the share of HBM left once weights, activations, and
CUDA-graph state are resident (Figure~\ref{fig:motivation}(a)).
Colocated, this saturation point is where TTFT starts to climb;
disaggregated, it caps a decode pool built from the same hardware at
the corresponding decode-only rate ($777$ tokens/s in our
measurements), no matter how much prefill capacity stands in front of
it. At $128$K, the same arithmetic admits only ${\sim}15$ requests.

\subsection{Decoupling Availability from Residency}
\label{sec:sparse-residency}

The capacity wall comes from tying together two requirements that need
not be identical. The model needs every past KV entry to be
\emph{logically available}, because a future top-$k$ selection may
refer to any position in the history. The GPU, however, only needs the
entries actually read by the current step. Full-HBM serving treats
``may be selected later'' as ``must stay in HBM now,'' which is simple
but overly strong. A serving system only needs to ensure that, once
the indexer emits $\mathcal{S}_t^{(\ell)}$, the selected KV entries
are on device before that layer's attention runs; entries outside
$\mathcal{S}_t^{(\ell)}$ can reside elsewhere without changing the
selected positions, attention scores, or outputs.

Decoupling alone, however, does not make offloading viable: the
CPU--GPU interconnect would immediately become the bottleneck. If
every layer's $k$ selections had to be fetched from host memory at
every step, a GLM-5.1 request would move roughly $200$\,MB of KV
records per generated token ($k{=}2048$ at ${\sim}100$\,KB of KV per
token summed across layers)---about $7$\,GB/s of sustained
host-to-device traffic per request at a TPOT of $30$\,ms, so a dozen
concurrent requests per GPU would saturate a PCIe Gen5 $\times$16 link
(${\sim}64$\,GB/s per direction) on misses alone. What closes this gap is locality in the selections themselves,
the structure noted in \S\ref{sec:intro}: consecutive decode steps
reselect largely overlapping positions~\citep{chen2025ess}, and
adjacent layers select correlated positions, which newer models make
explicit by sharing indexer outputs across
layers~\citep{bai2026indexcache,glm52_blog_2026}. On a LongBenchV2
selection trace, a modest LRU cache (twice the top-$k$ size) turns
$87\%$ of selections into device hits (\S\ref{sec:eval-cache}). This locality enables the
entire design: caching works because most selected records were
selected recently, and prefetching works because upcoming selections
are predictable---or, with shared indexers, known outright.

Realizing this separation without giving back its gains means bounding
each request's cache while preserving the locality above, hiding
residual miss latency behind computation, and doing both through an
interface that consumes only the emitted positions---the challenges
outlined in \S\ref{sec:intro}. The next section presents the design
that meets them; \S\ref{sec:eval} quantifies each mechanism.

\section{HiSparse Design}
\label{sec:design}

\subsection{Design Goals and Invariants}
\label{sec:design-goals}

\sys{} realizes the separation introduced in
\S\ref{sec:sparse-residency}: every KV entry remains logically
available to the sparse-attention algorithm, but only a bounded working
set is resident in GPU memory. The design is guided by the following
goals and invariants; Table~\ref{tab:notation} defines the notation
used throughout.

\begin{table}[t]
\centering
\caption{Notation.}
\label{tab:notation}
\small
\begin{tabular}{@{}ll@{}}
\toprule
Symbol & Meaning \\
\midrule
$L_{\text{ctx}}$ & context length of a request (tokens) \\
$k$ & tokens selected per query by the sparse indexer \\
$\mathcal{S}_t^{(\ell)}$ & selected set at decode step $t$, layer $\ell$; $|\mathcal{S}_t^{(\ell)}| = k$ \\
$B$ & GPU-cache capacity per request and layer (KV-record slots); $B \ge k$ \\
$N_{\text{batch}}$ & number of concurrent decode requests \\
$N_{\ell}$ & number of sparse-attention layers \\
$W_{\text{KV}}$ & KV elements stored per token per layer \\
$s$ & bytes per KV element \\
\bottomrule
\end{tabular}
\end{table}

\paragraphhead{Complete KV availability} For every active request,
\sys{} maintains a complete copy of the KV cache outside the decode
GPU's HBM. Any logical KV position selected by the sparse indexer can
therefore be recovered without recomputation.

\paragraphhead{Bounded device footprint} For each request and layer,
\sys{} reserves a fixed-size \emph{GPU cache} of $B$ logical KV-record
slots in HBM, which caches the KV records most recently selected for
that request and layer. $B$ is a serving configuration parameter with
$B \ge k$, independent of $L_{\text{ctx}}$. If the model stores
$W_{\text{KV}}$ elements per token per layer, the decode-side KV
footprint scales as
$N_{\text{batch}} N_{\ell} B W_{\text{KV}} s$ rather than
$N_{\text{batch}} N_{\ell} L_{\text{ctx}} W_{\text{KV}} s$, up to
metadata.

\paragraphhead{Exact sparse-attention outputs} Before a layer's
sparse-attention kernel runs, all KV entries in that layer's selected
set $\mathcal{S}_t^{(\ell)}$ must be materialized on device. \sys{}
may change where unselected KV entries reside, but it does not change
the selected positions, attention scores, or attention outputs.

\paragraphhead{Indexer-agnostic interface} \sys{} does not assume how
the selected set is produced. DSA, NSA, and Quest can use different
indexers; \sys{} only consumes the selected positions emitted for each request
and layer.

\paragraphhead{Miss latency off the critical path} Bounding residency
must not trade throughput for per-token latency. Selected-set misses
add work to every sparse-attention layer, so \sys{} treats their cost
as a first-class goal: cache management preserves the selection
locality of \S\ref{sec:sparse-residency} so that most selections hit
(\S\ref{sec:hierarchy}), resolving the remainder is a single fused
kernel launch (\S\ref{sec:kernel}), and prefetching overlaps
host-to-device transfers with computation in earlier layers
(\S\ref{sec:prefetch}).

\subsection{KV Hierarchy and Metadata}
\label{sec:hierarchy}

\begin{figure}[t]
\centering
\includegraphics[width=0.95\linewidth]{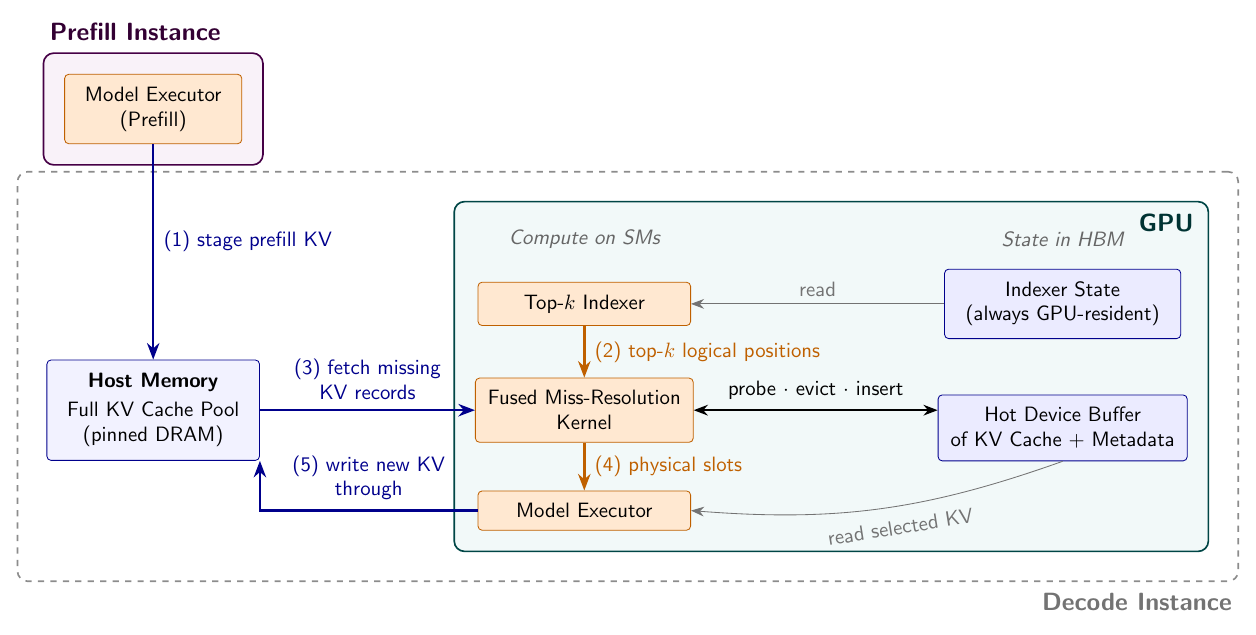}
\caption{\sys{} overview. Host memory keeps the authoritative full KV
cache of every active request; the GPU keeps only compact state
(indexer state, per-request-per-layer GPU caches with page-table and
LRU metadata) and runs all compute. (1)~Prefill writes each layer's KV
records to the host pool. During decode, (2)~the sparse indexer emits
selected logical positions; the fused \textsc{Resolve} kernel probes
the GPU cache, (3)~fetches missing records from the host pool while
evicting LRU victims, and (4)~hands physical cache slots to the
sparse-attention backend. (5)~KV records of newly generated tokens are
written through to the host pool.}
\label{fig:overview}
\end{figure}

\sys{} organizes KV state as a two-level hierarchy
(Figure~\ref{fig:overview}). The \textbf{host KV pool}, allocated in
pinned host DRAM, stores the authoritative full KV cache for active
requests. In colocated serving, prefill writes into the local host
pool. In disaggregated serving, prefill sends KV state to the decode
instance's host pool over the prefill-decode transfer path.

The decode GPU stores only a \textbf{GPU cache}---introduced in
\S\ref{sec:intro} and labeled ``hot device buffer'' in
Figures~\ref{fig:overview} and~\ref{fig:kernel}.
Conceptually, there is a $B$-slot cache for each request and layer;
each slot holds the layer-local KV record for one logical position.
The cache is more than a top-$k$ staging area: the selected entries
for the current step must be present before attention runs, while the
remaining $B-k$ slots keep recently useful records that are likely to
be hit by future selections. The number of resident records per
request and layer is thus at most $B$ and, once the cache warms,
typically above $k$; $B \ge k$ is required so that the current
selection always fits. A
\textbf{page table} maps each layer-local logical position either to a
slot in the cache or to a sentinel indicating that the record is
host-only. \textbf{LRU metadata} records the recency of resident slots
and directly drives replacement when selected entries miss. The
metadata is small relative to KV tensors, but every sparse layer must
consult and update it before its attention kernel can launch, so its
latency adds directly to every decode step; \sys{} therefore keeps it
GPU-resident and folds its updates into the fused kernel of
\S\ref{sec:kernel}.

Selection state, in contrast, never moves. The compact indexer state
identified in \S\ref{sec:sparse-attn}---DSA's per-token indexer keys,
NSA's compressed block keys, Quest's page summaries---remains
GPU-resident for the lifetime of a request, as do \sys{}'s page tables
and LRU metadata; \sys{} does not page any of them to host memory.
This always-resident state still grows with context length, but at a
per-token cost two to three orders of magnitude below the KV records
themselves: indexer keys and page-table entries together occupy at
most a few hundred bytes per token, versus ${\sim}100$\,KB of KV
records (\S\ref{sec:sparse-residency}). In our deployments it
amounts to hundreds of megabytes in total, small next to the multi-GB
per-request KV caches it replaces in HBM. Only attention KV records
travel through the hierarchy, so the indexer's own computation is
untouched.

\paragraphhead{Replacement policy} Each request-layer cache is managed
independently with LRU, with one deliberate refinement: within a step,
entries that \emph{hit} are promoted above the newly fetched misses in
the recency order. A record selected repeatedly across steps therefore
outranks a first-time selection, so under eviction pressure one-off
selections leave first and the cache accumulates the positions with
demonstrated multi-step reuse. \S\ref{sec:eval-cache} validates the
choice: recency is a good online proxy for future sparse selections,
tracking the trend of the offline B\'el\'ady optimum. Caches are
managed independently per layer: KV records are layer-local by
nature, and coordinating residency across layers adds little over
what LRU already captures (\S\ref{sec:eval-prefetch}); \sys{} instead
exploits cross-layer selection structure through prefetching
(\S\ref{sec:prefetch}).

\paragraphhead{Sizing the cache} $B$ trades capacity against latency.
Total decode KV HBM is $N_{\text{batch}} N_{\ell} B W_{\text{KV}} s$,
so a larger $B$ raises the hit rate but consumes HBM that could
otherwise admit more concurrent requests, and lengthens the resolve
kernel's metadata scans. In practice $B \in [2k, 4k]$ is the useful
range (\S\ref{sec:eval-micro}), and faster host-device links shift the
optimum toward $B{=}2k$ (\S\ref{sec:eval-gh200}).

\subsection{Request Lifecycle}
\label{sec:workflow}

A request moves through four phases, shown in
Figure~\ref{fig:overview}. The flow is identical in PD-colocated and
PD-disaggregated deployments; the two differ only in how prefill KV
reaches the host pool---written locally or sent over the
prefill--decode transfer path (\S\ref{sec:hierarchy}).

\paragraphhead{(1) Prefill and staging} The prefill engine processes
the prompt using the model's ordinary prefill path. As each layer's KV
state is produced, \sys{} writes it to the host KV pool. Compact
indexer state that must be consulted at every decode step, such as
DSA's lightning-indexer representation, remains on device.

\paragraphhead{(2) Admission} A request becomes schedulable for decode
once its host KV state is available and \sys{} has reserved its
per-layer GPU caches and metadata. The reserved KV capacity per
request is $N_{\ell} B W_{\text{KV}} s$---proportional to the
configured cache size $B$, a small multiple of $k$---rather than
$N_{\ell} L_{\text{ctx}} W_{\text{KV}} s$, so long-context requests no
longer consume decode HBM in proportion to their full history length
(for GLM-5.1 at $B{=}4096$, about $0.4$\,GB per request instead of
$13.09$\,GB at $128$K context---a ${\sim}30\times$ reduction).

\paragraphhead{(3) Layer decode} At decode step $t$ and layer $\ell$,
the indexer emits its selected set. The miss-resolution path
checks which selected logical positions are already resident, fetches
the missing layer-local KV records from the host pool, updates the page
table and LRU state, and emits a dense vector of physical device slots
aligned with the selected positions. The sparse-attention kernel for
that layer runs after this resolution completes. This describes the
synchronous path; \S\ref{sec:prefetch} overlaps part of the fetch work
with computation in preceding layers.

\paragraphhead{(4) Write-through of generated KV} Each newly generated
token's KV record is produced directly into a reserved slot of the
request's GPU cache, so the newest position is always resident for
upcoming selections. A dedicated backup stream then writes the record
through to the host pool, overlapped with the next step's computation
and ordered by events so that the backing copy is complete before any
later fetch can reference it.

\subsection{Fused Miss-Resolution Kernel}
\label{sec:kernel}

\begin{figure}[t]
\centering
\includegraphics[width=\linewidth]{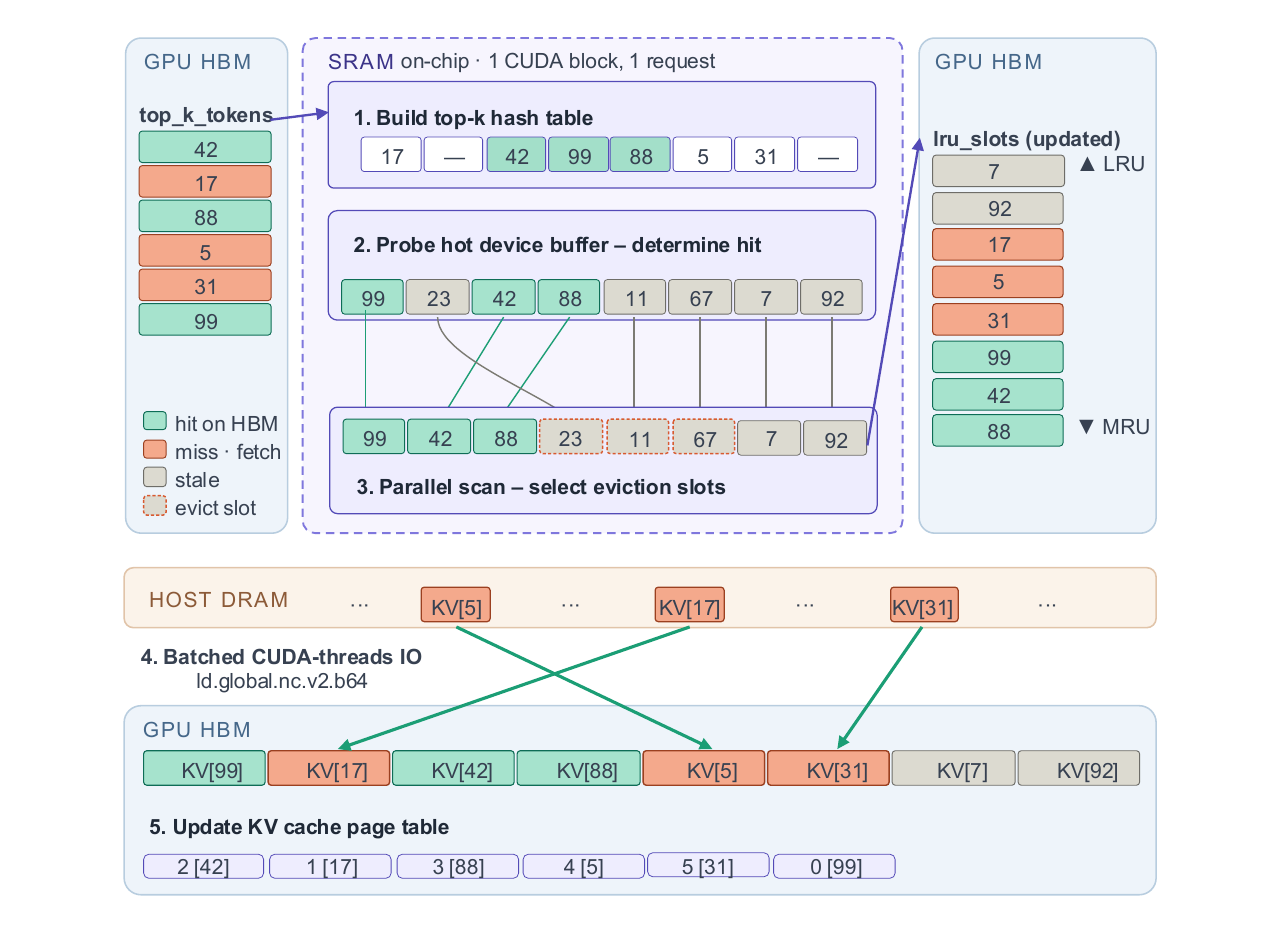}
\caption{The fused miss-resolution kernel. For one request and layer,
\textsc{Resolve} first builds a shared-memory hash table for the
selected logical positions. It then probes every GPU-cache slot
against the hash table in parallel, marks resident entries as hits or
eviction candidates, uses a parallel scan over those marks to choose
victims and update LRU metadata, fetches missing KV records from
pinned host memory, and emits physical device slots in top-$k$ order
for the sparse-attention backend.}
\label{fig:kernel}
\end{figure}

Miss resolution is on the critical path of every sparse-attention
layer. Given a layer's selected set, \sys{} must identify hits,
choose victims for misses, fetch missing KV records, update metadata,
and return physical device slots to the attention backend. Splitting
these operations across multiple CUDA launches would repeatedly
materialize transient state to HBM and add launch latency at every
layer. The intermediate state is also tightly coupled: hit/miss marks,
victim assignments, LRU updates, and the output device-location vector
all depend on the same selected set and resident-buffer metadata.
\sys{} therefore performs miss resolution with a single fused CUDA
kernel, \textsc{Resolve}, launched once per sparse layer with one CUDA
block handling each request's work item (Figure~\ref{fig:kernel}) and
captured inside SGLang's steady-state decode CUDA
graph~\citep{zheng2024sglang}, into which our implementation is
integrated. Because resolution consumes only the emitted logical
positions and \sys{}'s own metadata, this one kernel serves DSA, NSA,
and Quest alike---logical indices in, physical slots out, much like a
software-managed TLB.

\paragraphhead{Phase 1: stage selected positions} Threads
cooperatively load the selected positions into a shared-memory hash
table. This gives the rest of the kernel fast membership tests for the
selected set without repeatedly rereading the top-$k$ vector from HBM.

\paragraphhead{Phase 2: mark GPU-cache slots} As shown in
Figure~\ref{fig:kernel}, threads probe the hash table for the logical
position currently stored in each of the $B$ GPU-cache slots. The
kernel writes a compact mark for each slot: hit if the resident entry
is in the current selected set, and evictable otherwise.

\paragraphhead{Phase 3: scan marks and update LRU} The kernel then
performs a parallel scan over the per-slot marks. The scan compacts
evictable slots, selects enough victims for the missing selected
positions, and produces the updated LRU metadata for the post-step
buffer state. Hit slots are retained and promoted to the
most-recently-used end, victim slots are assigned to misses, and the
fetched misses are ordered just behind the hits, per the replacement
policy of \S\ref{sec:hierarchy}.

\paragraphhead{Phase 4: fetch missing KV records} Threads responsible
for misses copy the corresponding layer-local KV records from the
pinned host pool into their claimed device slots. \sys{} adopts the
GPU-assisted IO technique of Strata~\citep{xie2025strata}: instead of
staging DMA copies, GPU threads issue vectorized non-coherent loads
directly against pinned host memory (\texttt{ld.global.nc.v2.b64}),
which tolerates the scattered addresses of cache misses and reduces
transaction overhead on PCIe and NVLink-C2C
systems~\citep{nvidia_gracehopper_2026}. The per-thread transfer block
size is tuned so that fragmented miss reads still approach link
bandwidth.

\paragraphhead{Phase 5: publish attention inputs} Finally, the kernel
updates the page table and emits \texttt{top\_k\_device\_locs}, a dense
vector of physical device offsets aligned with the selected logical
positions. The downstream sparse-attention gather can then read the
selected KV records directly from the GPU cache.

\subsection{Layer-wise Prefetch}
\label{sec:prefetch}

Even with a fused resolve kernel, selected entries that miss in the
GPU cache may expose host-memory latency. \sys{} hides this cost by
prefetching across layers, targeting models that make cross-layer
selection locality explicit by reusing indexer outputs.
IndexCache~\citep{bai2026indexcache} partitions layers into what we
call \emph{anchor} layers, which run the top-$k$ indexer, and
\emph{shared} layers, which reuse the selection of their preceding
anchor; GLM-5.2 ships this design natively as IndexShare, sharing one
indexer across each group of four layers~\citep{glm52_blog_2026}. For
such models,
\sys{} does not need to guess: the moment an anchor layer emits its
selected set, the selected positions of every shared layer in its
group are known, several layers before their attention runs.

\paragraphhead{Exact prefetch with shared selections} \sys{}
exploits this with a \emph{plan-then-IO} scheme: the anchor's
\textsc{Resolve} additionally records its miss plan---which host
records move into which cache slots---and a copy-only kernel on a side
stream replays that plan into each shared layer's cache, overlapping
the transfers with the intervening layers' computation. Because every
shared layer's cache follows its anchor's slot layout in lockstep, a
shared layer reuses the anchor's slot table outright: it waits on its
prefetch-completion event and skips resolution entirely---no probing,
no LRU update, no synchronous host-memory load, and no speculative
traffic wasted. The copies themselves reuse the demand path's
GPU-assisted IO (\S\ref{sec:kernel}), so scattered prefetch reads run
near link bandwidth---which matters because prefetching shifts IO
earlier rather than eliminating it, drawing on the same host link
that serves demand misses (\S\ref{sec:eval-gh200}).

\paragraphhead{A speculative alternative} For models without shared
indices, we also explored a speculative variant that uses layer
$\ell$'s selected positions as a hint for layer $\ell{+}1$---adjacent
layers often select overlapping positions, and a wrong hint costs only
wasted transfers, never correctness. In our evaluation it produced
little end-to-end gain (\S\ref{sec:eval-prefetch}): the LRU-managed
GPU cache already captures most implicit cross-layer reuse, so hinted
records are usually resident, while the misses that remain are
precisely those the hints fail to predict. This negative result is
part of why we consider shared-index model co-design, rather than
deeper speculation, the right path to hiding miss latency.

\section{Evaluation}
\label{sec:eval}

Our evaluation asks four questions:
(1)~Does \sys{} improve end-to-end serving across context lengths,
models, sparse selectors, and hardware platforms?
(2)~Does the GPU cache exploit enough locality to keep miss
counts low?
(3)~What determines miss-resolution overhead and the choice of
GPU-cache size?
and (4)~how do faster host-device links and layer-wise prefetching
interact with emerging hardware and model-design trends?

\subsection{Setup}
\label{sec:eval-setup}

\paragraphhead{Models} We evaluate three sparse-attention families:
DeepSeek-V4-Flash, whose hybrid attention applies NSA-style top-$k$
selection over compressed KV entries (termed Compressed Sparse
Attention by DeepSeek)~\citep{deepseek_v4_2026,yuan2025nsa},
GLM-5.1-FP8 with DeepSeek Sparse
Attention (DSA)~\citep{glm5_2026,dsa2025,hisparse_blog_2026}, and
Qwen3-30B-A3B-Thinking-2507~\citep{qwen3_technical_report,qwen3_30b_a3b_thinking_2507}
with Quest as a training-free sparse
selector~\citep{tang2024quest}. The layer-wise prefetch study
(\S\ref{sec:eval-prefetch}) additionally uses GLM-5.2-FP8, which
shares DSA indexer selections across groups of layers
(IndexShare)~\citep{glm52_blog_2026,bai2026indexcache}. All experiments
select $k{=}2048$ tokens per query. For DeepSeek-V4-Flash, selection
operates at the granularity of $4$-token compressed KV entries, so its
top-$512$ selection covers $2048$ tokens; \sys{} manages these
compressed KV entries, which dominate the model's KV footprint, while
the remaining branch state stays GPU-resident. For the other models,
selection is token-level with $k{=}2048$; Quest requires no
architectural sparsity and serves as a training-free selector, applied
at every layer, over the ordinary dense-attention KV cache of the
smaller Qwen3 model. All
models serve with BF16 KV caches; the FP8 tag in model names refers to
weight precision.

\paragraphhead{Platforms} End-to-end serving experiments use the
platform indicated in each figure: DeepSeek-V4-Flash on $2\times$B200,
GLM-5.1-FP8 and GLM-5.2-FP8 on $8\times$H200, and Qwen3+Quest on a
GH200 node. The H200 node pairs its eight GPUs with $2$\,TB of host
DRAM; at the largest operating point (\S\ref{sec:eval-prefetch},
$256$ concurrent requests at $32$K input / $8$K output), the host KV
pool grows to roughly $1$\,TB of pinned host memory.

\paragraphhead{Baseline} All experiments compare against unmodified
SGLang v0.5.11 with the full KV cache resident in HBM, under an
otherwise identical deployment configuration (model, parallelism,
precision, and scheduler settings). We compare against no offloading
system: the closest efforts, ESS~\citep{chen2025ess} and
ECHO~\citep{liu2026echo}, are concurrent work---the former a prototype
evaluated in simulation, the latter specific to NSA
(\S\ref{sec:related}).

\paragraphhead{Workloads and metrics} We first evaluate end-to-end
serving with SGLang's standard \texttt{bench\_serving}
harness~\citep{zheng2024sglang,sglang_bench_serving_docs_2026}. For
each model, we sweep the input lengths shown in the corresponding
figures under fixed output lengths and closed-loop concurrency.
Concurrency sweeps (Figures~\ref{fig:motivation}, \ref{fig:v4-sweep},
and~\ref{fig:prefetch-results}) fix $32$K input / $8$K output: the
length is representative of long-context serving, and it is the regime
where the comparison is most informative---the full-KV baseline can
still admit a range of concurrencies at $32$K, so both systems trace
complete curves, whereas at much longer contexts the baseline
collapses to a few low-concurrency points and leaves little of the
trend to compare. Longer contexts, up to $200$K, are therefore
reported as peak-throughput comparisons
(Figure~\ref{fig:throughput-sweep}). These
runs report generation throughput (generated tokens/s), TTFT, and
TPOT. Generation throughput measures system output rate; TPOT measures
per-output-token latency after the first token. We then use ablations
to explain where the end-to-end gains come from: cache-policy
experiments replay sparse-selection traces from
LongBenchV2~\citep{bai2024longbenchv2} and kernel experiments report
miss-resolution time.

\subsection{End-to-End Benchmarks}
\label{sec:eval-end2end}

\begin{figure}[t]
\centering
\includegraphics[width=\linewidth]{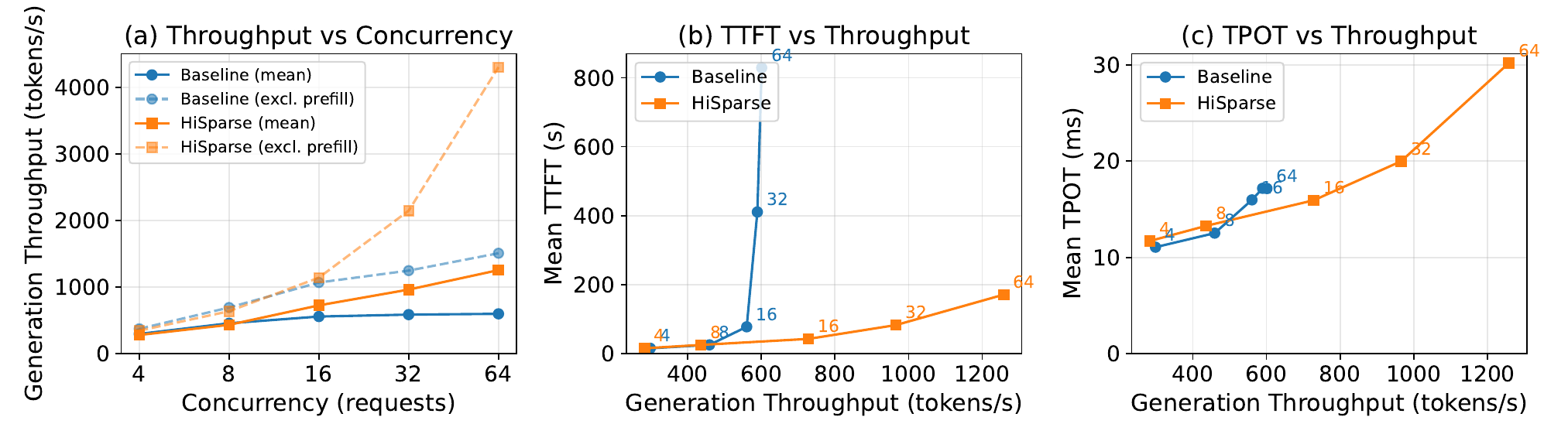}
\caption{End-to-end serving of DeepSeek-V4-Flash (NSA) on
$2\times$B200 in PD-colocated mode, $32$K input / $8$K output;
selection covers $2048$ tokens per step (top-$512$ over $4$-token
compressed KV entries).
\textbf{(a)}~Generation throughput vs.\ closed-loop concurrency
(solid: prefill+decode; dashed: decode-only, indicative of
PD-disaggregated decode-pool throughput); \textbf{(b)}~mean TTFT and
\textbf{(c)}~mean TPOT vs.\ achieved generation throughput.}
\label{fig:v4-sweep}
\end{figure}

\begin{figure}[t]
\centering
\includegraphics[width=\linewidth]{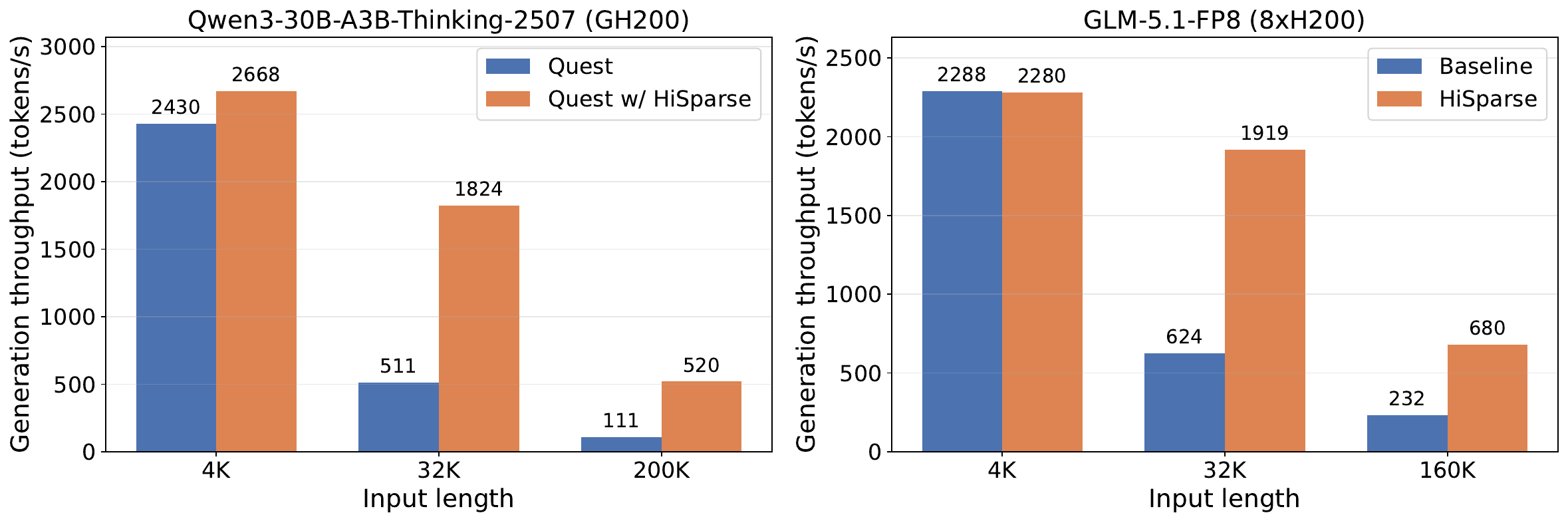}
\caption{Peak generation throughput across input lengths for two
additional sparse-attention families, both with $k{=}2048$ selected
tokens. \emph{Left:} Qwen3-30B-A3B with Quest on GH200.
\emph{Right:} GLM-5.1-FP8 with DSA on $8\times$H200.}
\label{fig:throughput-sweep}
\end{figure}

Figure~\ref{fig:v4-sweep} shows the end-to-end benchmark for
DeepSeek-V4-Flash at $32$K input / $8$K output. At low concurrency,
the baseline and \sys{} have similar throughput because the baseline
can still keep all active KV caches in HBM. Once concurrency rises,
the baseline saturates: additional requests cannot be admitted without
more KV capacity, so throughput remains nearly flat---the same wall
Figure~\ref{fig:motivation}(a) shows for GLM-5.1. \sys{} reduces
the per-request decode HBM footprint and continues to scale, lifting
generation throughput from $600$ to $1257$ tokens/s at concurrency
$64$ ($2.1\times$)---and from $1511$ to $4308$ tokens/s ($2.9\times$)
in decode-only terms. The gain is entirely a batch-size effect: \sys{} does not
make an individual decode step faster---it admits a larger decode
batch into the same HBM. At $32$K input the baseline can still hold a
few dozen requests, so the headroom here is moderate; with longer
inputs and outputs the baseline's feasible batch shrinks further and
the same effect becomes much more pronounced
(Figure~\ref{fig:throughput-sweep}).

The latency panels show why this matters for colocated serving. In the
baseline, decode saturation causes prefill chunks and new requests to
queue, so TTFT grows rapidly even though the per-token decode latency
has not collapsed: mean TTFT climbs from $26$\,s at concurrency $8$ to
$829$\,s at $64$, where \sys{} sits at $171$\,s. \sys{} leaves more
HBM headroom and drains decode work faster, keeping TTFT much lower at
higher achieved throughput. TPOT rises as \sys{} pushes into
higher-throughput operating points, but in the overlapping range it
remains comparable to the baseline ($15.9$ vs.\ $16.0$\,ms at
concurrency $16$), indicating that miss-resolution overhead is smaller
than the capacity benefit in the long-context regime.

Figure~\ref{fig:throughput-sweep} extends the result along two axes at
once: to other models, selectors, and platforms, and---most
importantly---to much longer contexts, sweeping input length up to
$200$K (Qwen3+Quest on GH200) and $160$K (GLM-5.1+DSA on H200). At $4$K, the full-KV baseline already fits a useful batch in
HBM, so \sys{} has little room to help: Qwen changes from $2430$ to
$2668$ tokens/s and GLM remains essentially unchanged ($2288$ vs.\
$2280$ tokens/s). At longer contexts, the baseline becomes
capacity-bound, while \sys{} keeps decode memory proportional to the
configured GPU-cache size. The resulting gains are large and grow with
context length: Qwen improves by $3.6\times$ at $32$K and $4.7\times$
at $200$K ($511$ to $1824$ and $111$ to $520$ tokens/s), while GLM
improves by $3.1\times$ at $32$K and $2.9\times$ at $160$K ($624$ to
$1919$ and $232$ to $680$ tokens/s). The GLM $32$K point summarizes the sweep already shown in
Figure~\ref{fig:motivation}, whose panels trace the full concurrency
and TTFT behavior at that operating point. This is the regime where
separating logical KV availability from GPU residency translates into
higher serving throughput. The decode-only curves carry the same
message for PD-disaggregated serving: with prefill time excluded, the
dashed curves in Figure~\ref{fig:v4-sweep}(a) estimate what a
dedicated decode pool could sustain. A full-KV decode pool is capped
by the batch its HBM can admit, and \sys{} raises exactly that
ceiling---the $2.9\times$ decode-only gain above. We do not run a
physically disaggregated deployment, due to our testbeds'
capacity; the decode-only rate serves as its proxy.

The capacity dividend is fungible: an operator who does not need more
throughput can take it as hardware savings instead. At matched batch
size, \sys{} shrinks the decode KV budget by the same multiplier that
otherwise raised concurrency---for GLM-5.1's ${\sim}60$-request batch
at $32$K, from ${\sim}240$\,GB of HBM to ${\sim}25$\,GB at $B{=}4096$,
a gap that widens with context ($13.09$ vs.\ $0.4$\,GB per request at
$128$K, \S\ref{sec:workflow}). In KV-dominated long-context
deployments, the same workload therefore fits in fewer GPUs or
cheaper, lower-HBM parts, at the cost of the per-token IO overhead of
\S\ref{sec:eval-prefetch}.

\subsection{GPU-Cache Locality and LRU}
\label{sec:eval-cache}

\begin{figure}[t]
\centering
\includegraphics[width=\linewidth]{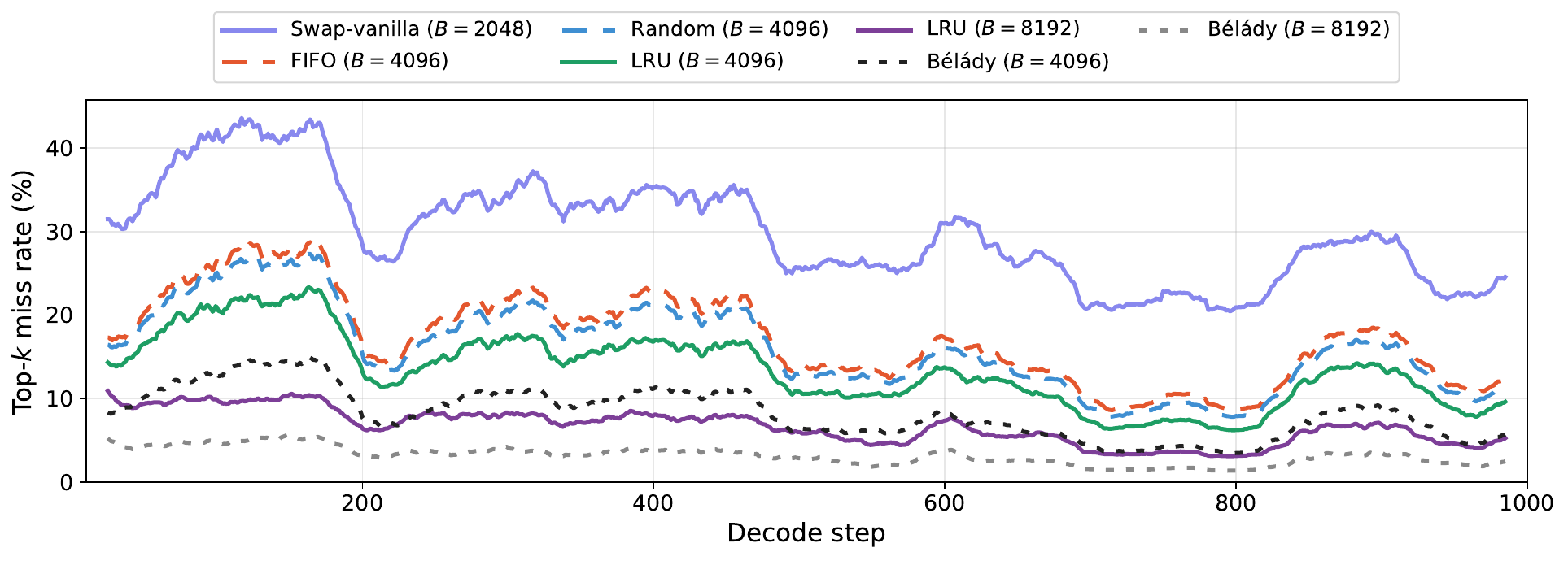}
\caption{Per-step top-$k$ miss rate (averaged across layers, smoothed)
when replaying the same LongBenchV2 sparse-selection trace of GLM-5.1
($k{=}2048$) under seven cache configurations; $B$ counts KV-record
slots per request and layer. Top-$k$-only staging
(\emph{Swap-vanilla}, $B{=}2048$) misses $30\%$ of each step's
selections on average because it retains no extra hot entries. With
the same $B{=}4096$ budget, LRU ($13.4\%$ mean) consistently
outperforms FIFO ($17.2\%$) and random ($16.1\%$) replacement and
follows the trend of the offline B\'el\'ady optimum ($8.2\%$).
Doubling the LRU cache to $B{=}8192$ halves the miss rate again
($6.7\%$), showing that retained locality translates directly into
fewer host-memory loads.}
\label{fig:miss-ablation}
\end{figure}

\sys{} deliberately uses the GPU cache as a hot KV cache, not merely
as a temporary top-$k$ staging area.
Figure~\ref{fig:miss-ablation} demonstrates why this choice matters
and why \sys{} manages the cache with LRU. The trace comes from a
GLM-5.1 request serving a $100{,}384$-token LongBenchV2 prompt and
decoding $1{,}799$ steps across all $78$ sparse layers; every policy
replays the identical per-layer top-$k$ selection stream (the figure
and the means below cover the first $1{,}000$ decode steps), so
differences reflect replacement decisions alone. Throughout, recall that $B$
counts KV-record slots \emph{per request and layer}
(Table~\ref{tab:notation}), so $B{=}4096$ is twice the $k{=}2048$
selection. Keeping exactly the current top-$k$ entries resident
($B{=}k$) misses $30\%$ of each step's selections on average, because
the selected set drifts across steps. Doubling the cache immediately
cuts misses, but the replacement policy determines how much locality
is preserved: at $B{=}4096$, LRU averages a $13.4\%$ miss rate,
consistently below FIFO ($17.2\%$) and random ($16.1\%$) replacement,
and tracks the shape of the offline-optimal B\'el\'ady
policy~\citep{belady1966}, indicating that recency is a good online
proxy for future sparse selections. This is also why the
miss-resolution kernel in \S\ref{sec:kernel} preserves resident hits
and updates LRU in place instead of rebuilding the cache from only the
current top-$k$ set. The same figure also shows what is left on the
table: doubling the LRU cache to $B{=}8192$ halves the miss rate to
$6.7\%$, while B\'el\'ady reaches $8.2\%$ already at $B{=}4096$ (and lower
still at $B{=}8192$, also shown), bounding what a smarter, predictive
replacement policy could still recover at the same capacity.

\subsection{Miss-Resolution Cost and Cache-Size Tradeoff}
\label{sec:eval-micro}

A low miss rate is only half the story: resolving the misses that
remain still costs time, and that cost shifts with cache size and
batch size. A larger GPU cache retains more locality but lengthens
metadata scans and consumes more HBM per request; at high batch size,
the misses of concurrent requests contend for host-memory bandwidth.
We therefore break the fused resolve kernel's time into phases, to
locate both the practical cache-size range and the bottleneck that
emerges once misses are rare.

\begin{figure}[t]
\centering
\includegraphics[width=\linewidth]{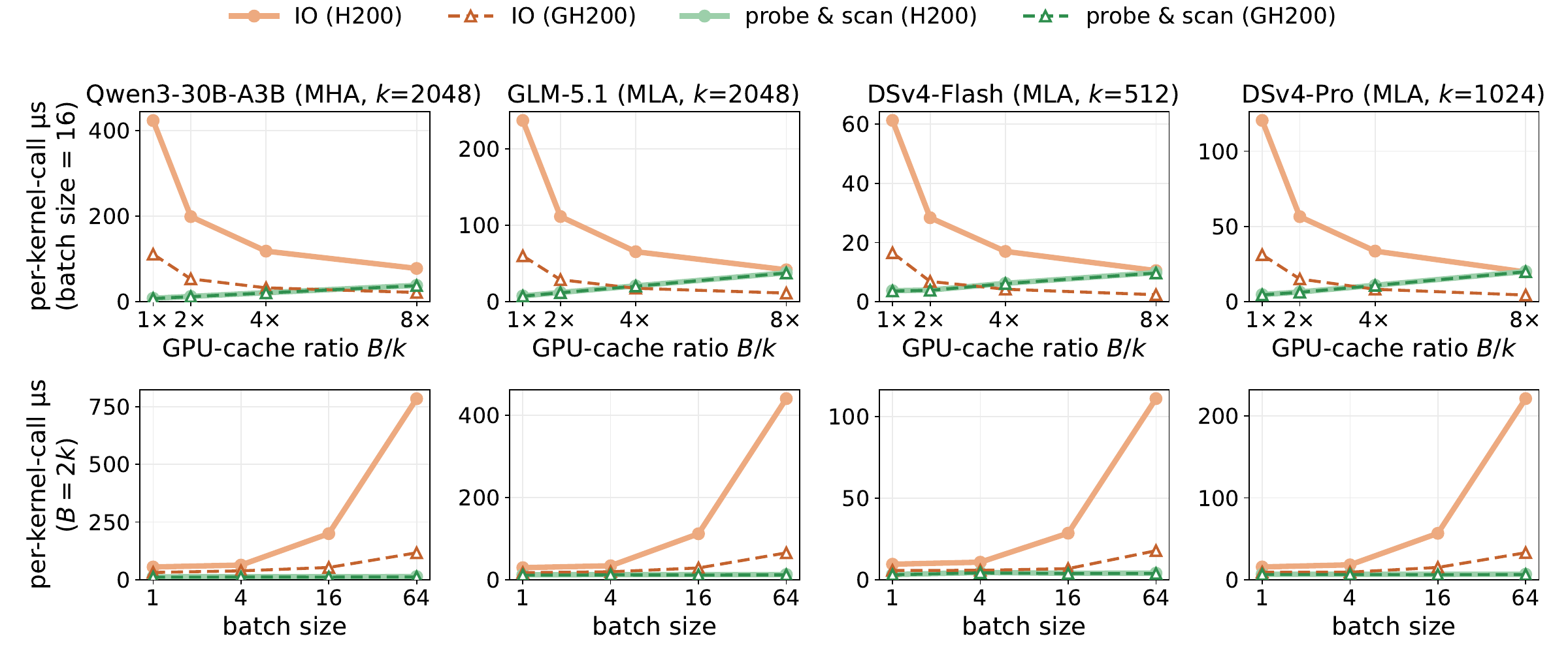}
\caption{Miss-resolution breakdown across models, GPU-cache sizes, and
platforms: H200 with a PCIe Gen5 host-device link (wide light lines,
filled markers) and GH200 with NVLink-C2C (dark dashed lines, open
markers). Panel titles give each model's top-$k$ in KV-record
granularity (\S\ref{sec:eval-setup}). \emph{IO} is the host-memory
fetch phase; \emph{probe \& scan} covers the metadata phases and is
nearly platform-independent---the dashed GH200 curve tracks the H200
band within ${\sim}5\%$; the residual phase (hash-table build and
output publication) is $1$--$4$\,\textmu s everywhere and omitted. Top row: per-kernel-call time at batch size
$16$ while varying the cache ratio $B/k$---larger caches reduce IO by
lowering misses but increase probe-and-scan work over resident slots.
Bottom row: per-kernel-call time at $B{=}2k$ while varying batch
size---at high batch, IO dominates on the PCIe link, while the faster
NVLink-C2C path compresses it.}
\label{fig:kernel-breakdown}
\end{figure}

Figure~\ref{fig:kernel-breakdown} shows this tuning tradeoff across
the three end-to-end models plus DeepSeek-V4-Pro (token-level
selection, $k{=}1024$), each at its native selection granularity. A
larger GPU cache reduces the number of
misses and therefore the IO portion of the resolve kernel. It is not
free: the kernel must probe and scan more resident slots, and a larger
cache also consumes more HBM per request---HBM that could otherwise
hold more concurrent requests, larger model weights, or more resident
MoE experts. Across the models in the figure, the useful region is
typically a small multiple of the selected set size, around
$2k$--$4k$. This range keeps enough hot entries to exploit locality
without turning metadata scans or HBM capacity into the new
bottleneck. For scale, the sparse-attention kernel itself takes
${\sim}60$\,\textmu s per layer in our GLM-5.2 decode profiles on H200
(per-GPU batch of $8$), so an unhidden $100$--$200$\,\textmu s resolve
would more than double a layer's attention critical path---which is
why keeping misses rare (\S\ref{sec:eval-cache}) and hiding the
remainder (\S\ref{sec:eval-prefetch}) both matter.

The bottom row isolates the batch-size effect at $B{=}2k$. The
probe-and-scan component is comparatively stable, while the IO
component grows quickly with batch size because more requests issue
miss loads concurrently. This result motivates the rest of the IO-side
design: making each fetch efficient with GPU-assisted IO and tuned
block sizes (\S\ref{sec:kernel}, \S\ref{sec:prefetch}), benefiting
from faster host-device links where the platform offers them
(\S\ref{sec:eval-gh200}), and hiding the remaining latency behind
computation with layer-wise prefetch (\S\ref{sec:eval-prefetch}).

\subsection{Bandwidth Sensitivity}
\label{sec:eval-gh200}

The previous experiment shows that host-memory IO can dominate
miss-resolution cost at high batch size. This raises a hardware-facing
question: if the CPU--GPU path becomes much faster, should \sys{} still
spend HBM on large GPU caches to avoid misses, or should it accept
more misses and reserve memory for concurrency? We study this question
on a GH200-class high-bandwidth host-device path~\citep{nvidia_gracehopper_2026}.

The dashed GH200 curves in Figure~\ref{fig:kernel-breakdown} repeat
the microbenchmark with only the platform changed and show that the
IO component compresses substantially---for GLM-5.1 at $B{=}2k$ and
batch size $16$, from $112$ to $29$\,\textmu s per call---so the
benefit of very large GPU caches diminishes: fetching a miss is
cheaper, while scanning a larger cache still costs time and consumes
HBM. This shifts the practical operating point toward smaller caches
such as
$B{=}2k$, which preserves more device memory for larger decode
batches. The result reinforces the central systems lesson: \sys{}
converts a capacity bottleneck into a tunable latency/bandwidth
problem, and platforms with faster host-device links make that
tradeoff more favorable.

In \sys{} this tuning is deliberately simple and static: $B$ is a
serving-configuration parameter fixed at deployment, chosen by
profiling sweeps like Figure~\ref{fig:kernel-breakdown} for the
target platform, with $B{=}2k$ a robust default. In our experiments
the preferred setting depended primarily on the platform's host-link
bandwidth (contrast the solid and dashed IO curves in
Figure~\ref{fig:kernel-breakdown}) rather than on the workload. Because caches are allocated per request,
admission-time or dynamic resizing is mechanically straightforward,
but we leave dynamic policies to future work. The next experiment
studies the complementary model-side opportunity: using cross-layer
selection locality to hide the remaining miss latency.

\subsection{Layer-wise Prefetching}
\label{sec:eval-prefetch}

Even on faster host-device links, a miss still lies on the critical
path unless its KV transfer can be overlapped with useful computation.
We evaluate \sys{}'s exact prefetch path (\S\ref{sec:prefetch}) on
GLM-5.2-FP8, which shares DSA indexer selections across layers via
IndexShare~\citep{glm52_blog_2026,bai2026indexcache}: of its $78$
layers, $21$ anchor layers run the indexer and the remaining $57$
layers reuse the selection of their preceding anchor. Once an anchor
layer emits its selected set, \sys{} issues the host-to-device loads
for the following shared layers' KV records and overlaps them with the
computation of the intervening layers. We compare four settings on
$8\times$H200 in PD-colocated mode at $32$K input / $8$K output with
closed-loop concurrency from $8$ to $256$: the full-KV baseline,
\sys{} with synchronous miss resolution (prefetch disabled), \sys{}
with exact prefetch, and a \emph{no-IO oracle}---\sys{} with the
resolve kernel's host-memory IO skipped entirely. The oracle serves
stale KV records on misses, so its outputs are invalid; because the
benchmark fixes output lengths, its timing remains a valid upper bound
on any IO-hiding scheme---stronger than perfectly overlapped
prefetching, which still places the same traffic on the host link. All
\sys{} configurations use $k{=}2048$ and $B{=}4096$ ($B{=}2k$), within
the $2k$--$4k$ range identified in \S\ref{sec:eval-micro}.

\begin{figure}[t]
\centering
\includegraphics[width=\linewidth]{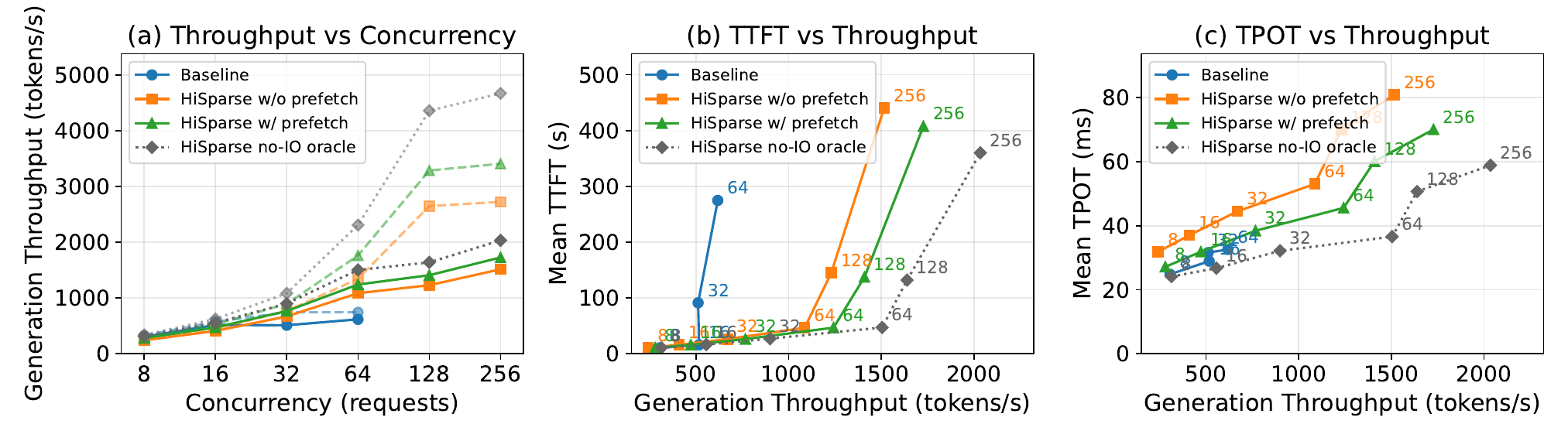}
\caption{Layer-wise exact prefetching with IndexCache-shared
selections: GLM-5.2-FP8 (DSA) on $8\times$H200 in PD-colocated mode,
$32$K input, $8$K output, $k{=}2048$, $B{=}4096$. The gray dotted
curve is the \emph{no-IO oracle}, which skips host-memory IO entirely
(outputs invalid; a performance bound on any IO-hiding scheme).
\textbf{(a)}~Generation throughput vs.\ closed-loop concurrency
(darker: prefill+decode; lighter: decode-only, indicative of
PD-disaggregated decode-pool throughput). The full-KV baseline
saturates once KV caches fill HBM; all \sys{} variants continue to
scale to $256$ requests.
\textbf{(b)}~Mean TTFT vs.\ achieved generation throughput; point
labels give the concurrency.
\textbf{(c)}~Mean TPOT vs.\ achieved generation throughput: the oracle
matches the baseline's TPOT at low concurrency---\sys{}'s entire
overhead is IO---and exact prefetch lowers TPOT by $13$--$15\%$ at
matched concurrency, closing roughly half of the gap between
synchronous resolution and the oracle.}
\label{fig:prefetch-results}
\end{figure}

Figure~\ref{fig:prefetch-results} shows the same capacity trend as
\S\ref{sec:eval-end2end}: the baseline saturates once full-context KV
caches fill HBM---mean TTFT jumps from $16$\,s at concurrency $16$ to
$91$\,s at $32$ and $275$\,s at $64$---while the \sys{} variants
scale to $256$ concurrent requests. Panel~(b) also shows that
prefill-side staging is effectively free: at low concurrency, where
TTFT contains no queueing, writing prefill KV through to the host pool
leaves mean TTFT unchanged ($10.7$\,s for the baseline vs.\
$10.7$--$10.8$\,s for all \sys{} variants at concurrency $8$).
Prefetching then recovers much of the remaining miss-resolution cost. At matched concurrency, exact
prefetch lowers mean TPOT by $13$--$15\%$ and raises generation
throughput by $14$--$17\%$ across the entire sweep, lifting peak
generation throughput from $618$ (baseline) and $1515$ (no prefetch)
to $1727$ tokens/s, a $2.8\times$ end-to-end gain over the full-KV
baseline. The no-IO oracle bounds what any IO-hiding scheme could
reach at $2034$ tokens/s: exact prefetch attains $85\%$ of this
ceiling, up from $74\%$ without prefetch. The decode-only curves give
the same bound for PD-disaggregated serving: the oracle's decode rate
reaches $4671$ tokens/s and exact prefetch $3410$ ($73\%$)---a wider
gap than in colocated mode, where prefill time dilutes the relative
cost of decode-side IO.

Panel~(c) quantifies the overhead against this explicit upper bound.
At low concurrency the oracle tracks the full-KV baseline's TPOT
($24.1$ vs.\ $24.8$\,ms at concurrency $8$), showing that the resolve
mechanism itself---hash probes, victim scans, LRU updates, and the KV
gather---adds no measurable per-token cost: \sys{}'s entire TPOT
overhead is host-memory IO. Against this bound, synchronous resolution
exposes $7.7$\,ms of IO per token at concurrency $8$ and $22.0$\,ms at
$256$; exact prefetch cuts the exposure to $3.0$\,ms and $11.2$\,ms
respectively, hiding roughly half of the IO overhead at matched
concurrency across the sweep. Much of the residual is structural:
anchor layers' own selections are not known in advance, so their
misses---$21$ of $78$ layers, roughly $27\%$ of the IO under a uniform
per-layer split---remain synchronous by construction. Discounting this
floor, prefetch hides two thirds to four fifths of the IO it can act
on; the remainder is overlap shortfall, since prefetching repositions
transfers rather than eliminating them and the traffic still contends
with demand misses on the same host link at high batch.

We also evaluated the speculative variant of \S\ref{sec:prefetch},
which uses layer $\ell$'s selection as a hint for layer $\ell{+}1$ in
models without shared indices. It raised the GPU-cache hit rate only
marginally and produced no measurable end-to-end gain: the hinted
positions are usually resident already, while the misses that
remain---positions newly entering the top-$k$---are precisely those
the hints cannot predict, so speculation adds host-link traffic
without removing critical-path loads. The takeaway is about
dependencies: overlapping miss IO with computation requires knowing a
layer's selections well before that layer runs, and speculation cannot
reliably supply them. Model co-design can---sharing selections across
layers removes the dependency outright and makes prefetch exact: the
difference between no measurable gain and the $13$--$15\%$ TPOT
reduction above.

\section{Discussion and Limitations}
\label{sec:limitations}

\paragraphhead{TPOT overhead from exposed IO} \sys{} trades HBM
capacity for host-memory traffic, which surfaces as a TPOT overhead:
$7$--$8$\,ms per token at matched low concurrency when misses are
resolved synchronously (\S\ref{sec:eval-prefetch}). The design layers
mitigations---LRU caching (\S\ref{sec:eval-cache}), GPU-assisted IO
(\S\ref{sec:kernel}), faster links (\S\ref{sec:eval-gh200})---and
model co-design hides roughly half of what remains: with shared
selections, exact prefetch cuts the exposed IO to about $3$\,ms per
token at low concurrency, against a no-IO oracle showing the resolve
mechanism itself costs nothing (\S\ref{sec:eval-prefetch}). When serving is not
capacity-bound---short contexts or low concurrency---\sys{} offers no
benefit to offset this overhead and can simply be disabled.

\paragraphhead{Beyond throughput: contexts that do not fit in HBM}
The capacity benefit is not only a batch-size multiplier. Under
full-KV serving, a request whose KV cache exceeds free HBM cannot be
admitted at \emph{any} concurrency---device memory, not the model's
context window, caps the servable context (a $1$M-token GLM-5.1
request needs over $100$\,GB of KV, \S\ref{sec:intro}). Because
\sys{} bounds decode-side HBM by the GPU-cache size regardless of
context length, the same mechanism that converts capacity into batch
size at moderate contexts converts it into feasibility at extreme
ones: the maximum servable context is set by host-tier capacity
instead. This follows directly from the footprint bound
(\S\ref{sec:workflow}), so we do not evaluate it separately.

\paragraphhead{Host-memory capacity} The more fundamental limitation
today is the size of the second tier, not the overhead. \sys{}
presumes host DRAM is much larger than HBM---true on terabyte-class
PCIe H200 servers, but not on Grace-based GB200/GB300 systems, where
each Grace CPU's ${\sim}480$\,GB of LPDDR is comparable to (for GB300,
smaller than) the paired GPUs' aggregate HBM. When the second tier is
no bigger than the first, the capacity multiplier that \sys{} converts
into batch size shrinks. NVMe or network-attached tiers recover
capacity at higher latency, raising the stakes for locality and
overlap.

\paragraphhead{Co-design implications} We hope this exploration
informs future co-design. Models that reserve headroom for
overlap---sharing selections across layers or emitting them
early~\citep{bai2026indexcache,glm52_blog_2026}---turn KV placement
from a latency problem into a scheduling one; platforms with larger
host memory and faster CPU--GPU interconnects convert directly into
smaller GPU caches and larger batches (\S\ref{sec:eval-gh200}).
Sparse attention makes per-step KV demand small and predictable;
modest accommodations in models and hardware would compound what
\sys{} already exploits today.

\section{Related Work}
\label{sec:related}

\paragraph{LLM serving and KV memory management.}
PagedAttention and vLLM manage \kv{} memory in fixed-size pages to
reduce fragmentation and enable higher serving throughput~\citep{kwon2023pagedattention}.
SGLang provides a high-throughput serving runtime with KV-reuse
optimizations for structured LLM programs~\citep{zheng2024sglang}.
Serving systems also split or coordinate prefill and decode because
the two phases stress different
resources~\citep{splitwise2024,distserve2024}, and IO-aware attention
kernels reduce compute-side memory traffic without shrinking the
resident \kv{}
footprint~\citep{dao2022flashattention,dao2024flashattention2}. \sys{}
is complementary to all of these: it reduces the decode-side \hbm{}
footprint of each long-context sparse-attention request, which helps
both disaggregated decode pools and colocated prefill/decode
deployments.

\paragraph{KV compression and eviction.}
A large body of work shrinks the \kv{} cache by discarding or
approximating entries: eviction policies such as H2O, StreamingLLM,
and SnapKV permanently drop tokens deemed
unimportant~\citep{h2o2023,streamingllm2024,snapkv2024}, and
quantization reduces bytes per entry~\citep{kivi2024}. These methods
trade output fidelity for capacity. \sys{} is orthogonal: it relocates
\kv{} entries rather than discarding or approximating them, leaving
sparse-attention outputs unchanged.

\paragraph{Hierarchical and offloaded KV caches.}
Systems such as FlexGen, HiCache, CachedAttention, and Mooncake place
model or \kv{} state across \gpu{}, host, and storage tiers for
throughput-oriented or cross-turn
serving~\citep{flexgen2023,hicache_blog_2025,attentionstore2024,mooncake2025};
\sys{} borrows Strata's GPU-assisted IO technique for its host-memory
fetch path~\citep{xie2025strata}. Closer to \sys{}, several systems
offload \kv{} state to host memory and fetch or use it selectively
during decode: InfiniGen speculatively prefetches entries predicted to
be important~\citep{lee2024infinigen}, ShadowKV selects via low-rank
keys kept on \gpu{} and offloads values~\citep{sun2025shadowkv},
MagicPIG samples \kv{} entries with CPU-resident
LSH~\citep{chen2025magicpig}, ArkVale evicts cold pages but recalls
them by page summaries~\citep{chen2024arkvale}, and PQCache retrieves
entries via product quantization~\citep{zhang2025pqcache}; NEO and
FastDecode instead offload attention computation itself to the
CPU~\citep{jiang2025neo,he2024fastdecode}. The retrieval-based systems
graft their own approximate selector onto dense-attention models, so
outputs deviate from the underlying model; the CPU-compute systems
remain exact but place attention on the \cpu{}. \sys{} instead serves
the model's \emph{own} top-$k$ selections---exact by
construction---and keeps all compute on the \gpu{}. Closest to \sys{}
are two concurrent systems. The ESS prototype offloads
DeepSeek-V3.2's latent cache to host memory behind a \gpu{} hot
cache, with an architecture specialized to that model and evaluated
in simulation~\citep{chen2025ess}. ECHO offloads KV state for NSA
models and centers on prefetching, whose benefit hinges on accurately
predicting upcoming selections~\citep{liu2026echo}. \sys{} spans
trained and training-free selectors (DSA, NSA, Quest) through an
indexer-agnostic interface, puts locality first---the LRU-managed
cache keeps most selections resident, reducing IO load rather than
overlapping it---resolves misses with a single fused kernel captured
in the decode CUDA graph, and reserves prefetching for models whose
shared selections make it exact rather than predictive.

\paragraph{Sparse attention.}
Early architectures constrain attention with fixed patterns,
content-based bucketing, or low-rank
projection~\citep{longformer,reformer,linformer}. Trained query-dependent
sparse-attention architectures such as DSA/NSA and hybrid attention
models reduce attention compute and active \kv{}
reads~\citep{dsa2025,glm5_2026,yuan2025nsa,deepseek_v4_2026}, and
training-free methods such as Quest select query-dependent \kv{} pages
at inference time~\citep{tang2024quest}. \sys{} is not a new
sparse-attention algorithm; it is a serving-system layer that makes
these algorithms' inactive state cheap in \hbm{} capacity.

\section{Conclusion}

Top-$k$ sparse attention leaves serving systems paying full-context \hbm{} rent for \kv{} entries that each decode step barely reads. \sys{} ends this mismatch by decoupling logical availability from physical residency: each request's complete \kv{} history lives in host memory, its decode footprint is bounded by a small \gpu{} cache whose LRU management turns selection locality into hits, a fused kernel resolves each layer's selections inside the decode CUDA graph, and models that share selections across layers make prefetching exact, overlapping much of the remaining IO with computation. Because only \kv{} placement changes, model outputs are unchanged. Across DSA, NSA, and Quest on H200, B200, and GH200 platforms, \sys{} improves peak long-context generation throughput by up to $4.7\times$ at comparable per-token latency; a no-IO oracle shows the resolution mechanism itself adds no measurable cost and that exact prefetch reaches $85\%$ of the throughput any IO-hiding scheme could attain.

Beyond throughput, bounding decode-side \hbm{} extends the servable context past what \hbm{} alone can hold, with the ceiling set by host-tier capacity instead. \sys{} ships in upstream SGLang, and its broader lesson is one of co-design: sparse attention makes per-step \kv{} demand small and predictable, and modest accommodations in models and hardware---selections shared or announced early, faster host--device links---compound what hierarchical \kv{} placement already delivers today.

\section*{Acknowledgements}
We would like to thank the Alibaba Cloud TairKVCache team, the Ant
Group SCT Inference team, the Baidu Baige AI team and Zhipu AI for 
their open-source contributions to \sys{}. We are grateful to 
Shangming Cai, Teng Ma, and Xingyu Ling from Alibaba Cloud 
for their constructive feedback. This research was supported
in part by the Stanford Platform Lab and its affiliates. Zhiqiang Xie
was supported by the NVIDIA Graduate Fellowship. We thank RadixArk for
providing computational resources.

\bibliographystyle{plainnat}
\bibliography{hisparse_refs}

\begin{thebibliography}{43}
\providecommand{\natexlab}[1]{#1}
\providecommand{\url}[1]{\texttt{#1}}
\expandafter\ifx\csname urlstyle\endcsname\relax
  \providecommand{\doi}[1]{doi: #1}\else
  \providecommand{\doi}{doi: \begingroup \urlstyle{rm}\Url}\fi

\bibitem[Bai et~al.(2025)Bai, Tu, Zhang, Peng, Wang, Lv, Cao, Xu, Hou, Dong,
  Tang, and Li]{bai2024longbenchv2}
Yushi Bai, Shangqing Tu, Jiajie Zhang, Hao Peng, Xiaozhi Wang, Xin Lv, Shulin
  Cao, Jiazheng Xu, Lei Hou, Yuxiao Dong, Jie Tang, and Juanzi Li.
\newblock {LongBench} v2: Towards deeper understanding and reasoning on
  realistic long-context multitasks.
\newblock In \emph{Proceedings of the 63rd Annual Meeting of the Association
  for Computational Linguistics (Volume 1: Long Papers)}, pages 3639--3664,
  2025.
\newblock \doi{10.18653/v1/2025.acl-long.183}.
\newblock URL \url{https://aclanthology.org/2025.acl-long.183/}.

\bibitem[Bai et~al.(2026)Bai, Dong, Jiang, Lv, Du, Zeng, Tang, and
  Li]{bai2026indexcache}
Yushi Bai, Qian Dong, Ting Jiang, Xin Lv, Zhengxiao Du, Aohan Zeng, Jie Tang,
  and Juanzi Li.
\newblock {IndexCache}: Accelerating sparse attention via cross-layer index
  reuse, 2026.
\newblock URL \url{https://arxiv.org/abs/2603.12201}.

\bibitem[Belady(1966)]{belady1966}
Laszlo~A. Belady.
\newblock A study of replacement algorithms for a virtual-storage computer.
\newblock \emph{IBM Systems Journal}, 5\penalty0 (2):\penalty0 78--101, 1966.
\newblock \doi{10.1147/sj.52.0078}.

\bibitem[Beltagy et~al.(2020)Beltagy, Peters, and Cohan]{longformer}
Iz~Beltagy, Matthew~E. Peters, and Arman Cohan.
\newblock Longformer: The long-document transformer, 2020.
\newblock URL \url{https://arxiv.org/abs/2004.05150}.

\bibitem[Chen et~al.(2024)Chen, Wang, Cao, Wu, Zheng, Li, Wei, Yan, Li, and
  Liang]{chen2024arkvale}
Renze Chen, Zhuofeng Wang, Beiquan Cao, Tong Wu, Size Zheng, Xiuhong Li,
  Xuechao Wei, Shengen Yan, Meng Li, and Yun Liang.
\newblock {ArkVale}: Efficient generative {LLM} inference with recallable
  key-value eviction.
\newblock In \emph{Advances in Neural Information Processing Systems}, 2024.

\bibitem[Chen et~al.(2025{\natexlab{a}})Chen, Zhang, He, Liu, Zhang, Zhou, Li,
  Zeng, Li, Qian, Li, and Li]{chen2025ess}
Xinhang Chen, Chao Zhang, Jiahuan He, Wei Liu, Jianming Zhang, Wenlong Zhou,
  Xiao Li, Pai Zeng, Shiyong Li, Yuanpan Qian, Dong Li, and Zhaogeng Li.
\newblock {ESS}: An offload-centric latent-cache management architecture for
  {DeepSeek-V3.2-Exp}, 2025{\natexlab{a}}.
\newblock URL \url{https://arxiv.org/abs/2512.10576}.

\bibitem[Chen et~al.(2025{\natexlab{b}})Chen, Sadhukhan, Ye, Zhou, Zhang,
  Nolte, Tian, Douze, Bottou, Jia, and Chen]{chen2025magicpig}
Zhuoming Chen, Ranajoy Sadhukhan, Zihao Ye, Yang Zhou, Jianyu Zhang, Niklas
  Nolte, Yuandong Tian, Matthijs Douze, Leon Bottou, Zhihao Jia, and Beidi
  Chen.
\newblock {MagicPIG}: {LSH} sampling for efficient {LLM} generation.
\newblock In \emph{International Conference on Learning Representations},
  2025{\natexlab{b}}.

\bibitem[Dao(2024)]{dao2024flashattention2}
Tri Dao.
\newblock {FlashAttention-2}: Faster attention with better parallelism and work
  partitioning.
\newblock In \emph{International Conference on Learning Representations}, 2024.
\newblock URL \url{https://openreview.net/forum?id=mZn2Xyh9Ec}.

\bibitem[Dao et~al.(2022)Dao, Fu, Ermon, Rudra, and
  R{\'e}]{dao2022flashattention}
Tri Dao, Daniel~Y. Fu, Stefano Ermon, Atri Rudra, and Christopher R{\'e}.
\newblock {FlashAttention}: Fast and memory-efficient exact attention with
  {IO}-awareness.
\newblock In \emph{Advances in Neural Information Processing Systems}, 2022.
\newblock URL \url{https://openreview.net/forum?id=H4DqfPSibmx}.

\bibitem[{DeepSeek-AI}(2025{\natexlab{a}})]{deepseek_v32_hf_2025}
{DeepSeek-AI}.
\newblock {DeepSeek-V3.2}: Efficient reasoning \& agentic {AI}.
\newblock Hugging Face model card, 2025{\natexlab{a}}.
\newblock URL \url{https://huggingface.co/deepseek-ai/DeepSeek-V3.2}.
\newblock Accessed 2026-05-04.

\bibitem[{DeepSeek-AI}(2025{\natexlab{b}})]{dsa2025}
{DeepSeek-AI}.
\newblock {DeepSeek-V3.2}: Pushing the frontier of open large language models,
  2025{\natexlab{b}}.
\newblock URL \url{https://arxiv.org/abs/2512.02556}.

\bibitem[{DeepSeek-AI}(2026)]{deepseek_v4_2026}
{DeepSeek-AI}.
\newblock {DeepSeek-V4}: Towards highly efficient million-token context
  intelligence, 2026.
\newblock URL \url{https://arxiv.org/abs/2606.19348}.

\bibitem[Gao et~al.(2024)Gao, He, Sharma, Kang, Jevdjic, Deng, Yang, Yu, and
  Zuo]{attentionstore2024}
Bin Gao, Zhuomin He, Puru Sharma, Qingxuan Kang, Djordje Jevdjic, Junbo Deng,
  Xingkun Yang, Zhou Yu, and Pengfei Zuo.
\newblock Cost-efficient large language model serving for multi-turn
  conversations with {CachedAttention}.
\newblock In \emph{USENIX Annual Technical Conference (ATC)}, 2024.
\newblock URL
  \url{https://www.usenix.org/conference/atc24/presentation/gao-bin-cost}.

\bibitem[{GLM-5 Team}(2026)]{glm5_2026}
{GLM-5 Team}.
\newblock {GLM-5}: from vibe coding to agentic engineering, 2026.
\newblock URL \url{https://arxiv.org/abs/2602.15763}.

\bibitem[He and Zhai(2024)]{he2024fastdecode}
Jiaao He and Jidong Zhai.
\newblock {FastDecode}: High-throughput {GPU}-efficient {LLM} serving using
  heterogeneous pipelines, 2024.
\newblock URL \url{https://arxiv.org/abs/2403.11421}.

\bibitem[Jiang et~al.(2025)Jiang, Zhou, Cao, Stoica, and Yu]{jiang2025neo}
Xuanlin Jiang, Yang Zhou, Shiyi Cao, Ion Stoica, and Minlan Yu.
\newblock {NEO}: Saving {GPU} memory crisis with {CPU} offloading for online
  {LLM} inference.
\newblock In \emph{Proceedings of Machine Learning and Systems (MLSys)}, 2025.

\bibitem[Kitaev et~al.(2020)Kitaev, Kaiser, and Levskaya]{reformer}
Nikita Kitaev, Lukasz Kaiser, and Anselm Levskaya.
\newblock Reformer: The efficient transformer.
\newblock In \emph{International Conference on Learning Representations}, 2020.

\bibitem[Kwon et~al.(2023)Kwon, Li, Zhuang, Sheng, Zheng, Yu, Gonzalez, Zhang,
  and Stoica]{kwon2023pagedattention}
Woosuk Kwon, Zhuohan Li, Siyuan Zhuang, Ying Sheng, Lianmin Zheng, Cody~Hao Yu,
  Joseph~E. Gonzalez, Hao Zhang, and Ion Stoica.
\newblock Efficient memory management for large language model serving with
  {PagedAttention}.
\newblock In \emph{Proceedings of the 29th ACM Symposium on Operating Systems
  Principles (SOSP)}, pages 611--626, 2023.
\newblock \doi{10.1145/3600006.3613165}.
\newblock URL \url{https://arxiv.org/abs/2309.06180}.

\bibitem[Lee et~al.(2024)Lee, Lee, Seo, and Sim]{lee2024infinigen}
Wonbeom Lee, Jungi Lee, Junghwan Seo, and Jaewoong Sim.
\newblock {InfiniGen}: Efficient generative inference of large language models
  with dynamic {KV} cache management.
\newblock In \emph{18th USENIX Symposium on Operating Systems Design and
  Implementation (OSDI)}, 2024.

\bibitem[Li et~al.(2024)Li, Huang, Yang, Venkitesh, Locatelli, Ye, Cai, Lewis,
  and Chen]{snapkv2024}
Yuhong Li, Yingbing Huang, Bowen Yang, Bharat Venkitesh, Acyr Locatelli,
  Hanchen Ye, Tianle Cai, Patrick Lewis, and Deming Chen.
\newblock {SnapKV}: {LLM} knows what you are looking for before generation.
\newblock In \emph{Advances in Neural Information Processing Systems}, 2024.

\bibitem[Liu et~al.(2026)Liu, Chen, Li, Ning, Lin, Yao, Chen, Sun, Zhao, and
  Guo]{liu2026echo}
Guangda Liu, Wenhao Chen, Chengwei Li, Zhenyu Ning, Jing Lin, Yiwu Yao, Quan
  Chen, Shixuan Sun, Jieru Zhao, and Minyi Guo.
\newblock {ECHO}: Efficient {KV} cache offloading with lossless prefetching for
  serving native sparse attention {LLMs}.
\newblock In \emph{20th USENIX Symposium on Operating Systems Design and
  Implementation (OSDI)}, 2026.

\bibitem[Liu et~al.(2024)Liu, Yuan, Jin, Zhong, Xu, Braverman, Chen, and
  Hu]{kivi2024}
Zirui Liu, Jiayi Yuan, Hongye Jin, Shaochen Zhong, Zhaozhuo Xu, Vladimir
  Braverman, Beidi Chen, and Xia Hu.
\newblock {KIVI}: A tuning-free asymmetric 2bit quantization for {KV} cache.
\newblock In \emph{International Conference on Machine Learning}, pages
  32332--32344, 2024.

\bibitem[{NVIDIA}(2026)]{nvidia_gracehopper_2026}
{NVIDIA}.
\newblock {NVIDIA} {GH200} {Grace Hopper} superchip.
\newblock Product page, 2026.
\newblock URL
  \url{https://www.nvidia.com/en-us/data-center/grace-hopper-superchip/}.
\newblock Accessed 2026-05-04.

\bibitem[Patel et~al.(2024)Patel, Choukse, Zhang, Shah, Goiri, Maleki, and
  Bianchini]{splitwise2024}
Pratyush Patel, Esha Choukse, Chaojie Zhang, Aashaka Shah, {\'I}{\~n}igo Goiri,
  Saeed Maleki, and Ricardo Bianchini.
\newblock Splitwise: Efficient generative {LLM} inference using phase
  splitting.
\newblock In \emph{Proceedings of the 51st Annual International Symposium on
  Computer Architecture (ISCA)}, 2024.
\newblock \doi{10.1109/ISCA59077.2024.00019}.

\bibitem[Qin et~al.(2025)Qin, Li, He, Cui, Ren, Zhang, Wu, Zheng, and
  Xu]{mooncake2025}
Ruoyu Qin, Zheming Li, Weiran He, Jialei Cui, Feng Ren, Mingxing Zhang, Yongwei
  Wu, Weimin Zheng, and Xinran Xu.
\newblock Mooncake: Trading more storage for less computation---a
  {KVCache}-centric architecture for serving {LLM} chatbot.
\newblock In \emph{23rd USENIX Conference on File and Storage Technologies
  (FAST)}, 2025.

\bibitem[{Qwen Team}(2025)]{qwen3_30b_a3b_thinking_2507}
{Qwen Team}.
\newblock {Qwen3-30B-A3B-Thinking-2507}.
\newblock Hugging Face model card, 2025.
\newblock URL \url{https://huggingface.co/Qwen/Qwen3-30B-A3B-Thinking-2507}.
\newblock Accessed 2026-06-18.

\bibitem[{SGLang Project}(2026{\natexlab{a}})]{sglang_bench_serving_docs_2026}
{SGLang Project}.
\newblock Bench serving guide.
\newblock SGLang Documentation, 2026{\natexlab{a}}.
\newblock URL \url{https://docs.sglang.io/docs/developer_guide/bench_serving}.
\newblock Accessed 2026-06-18.

\bibitem[{SGLang Project}(2026{\natexlab{b}})]{sglang_hisparse_docs_2026}
{SGLang Project}.
\newblock {HiSparse}: Hierarchical sparse attention.
\newblock SGLang Documentation, 2026{\natexlab{b}}.
\newblock URL
  \url{https://docs.sglang.io/docs/advanced_features/hisparse_guide}.
\newblock Accessed 2026-05-04.

\bibitem[Sheng et~al.(2023)Sheng, Zheng, Yuan, Li, Ryabinin, Chen, Liang,
  R{\'e}, Stoica, and Zhang]{flexgen2023}
Ying Sheng, Lianmin Zheng, Binhang Yuan, Zhuohan Li, Max Ryabinin, Beidi Chen,
  Percy Liang, Christopher R{\'e}, Ion Stoica, and Ce~Zhang.
\newblock {FlexGen}: High-throughput generative inference of large language
  models with a single {GPU}.
\newblock In \emph{International Conference on Machine Learning}, pages
  31094--31116, 2023.
\newblock URL \url{https://proceedings.mlr.press/v202/sheng23a.html}.

\bibitem[Sun et~al.(2025)Sun, Chang, Bao, Zheng, Zheng, Liu, Dong, Chi, and
  Chen]{sun2025shadowkv}
Hanshi Sun, Li-Wen Chang, Wenlei Bao, Size Zheng, Ningxin Zheng, Xin Liu, Harry
  Dong, Yuejie Chi, and Beidi Chen.
\newblock {ShadowKV}: {KV} cache in shadows for high-throughput long-context
  {LLM} inference.
\newblock In \emph{International Conference on Machine Learning}, pages
  57355--57373, 2025.

\bibitem[Tang et~al.(2024)Tang, Zhao, Zhu, Xiao, Kasikci, and
  Han]{tang2024quest}
Jiaming Tang, Yilong Zhao, Kan Zhu, Guangxuan Xiao, Baris Kasikci, and Song
  Han.
\newblock Quest: Query-aware sparsity for efficient long-context {LLM}
  inference.
\newblock In \emph{International Conference on Machine Learning}, pages
  47901--47911, 2024.
\newblock URL \url{https://proceedings.mlr.press/v235/tang24l.html}.

\bibitem[Wang et~al.(2020)Wang, Li, Khabsa, Fang, and Ma]{linformer}
Sinong Wang, Belinda~Z. Li, Madian Khabsa, Han Fang, and Hao Ma.
\newblock Linformer: Self-attention with linear complexity, 2020.
\newblock URL \url{https://arxiv.org/abs/2006.04768}.

\bibitem[Xiao et~al.(2024)Xiao, Tian, Chen, Han, and Lewis]{streamingllm2024}
Guangxuan Xiao, Yuandong Tian, Beidi Chen, Song Han, and Mike Lewis.
\newblock Efficient streaming language models with attention sinks.
\newblock In \emph{International Conference on Learning Representations}, 2024.

\bibitem[Xie(2025)]{hicache_blog_2025}
Zhiqiang Xie.
\newblock {SGLang} {HiCache}: Fast hierarchical {KV} caching with your favorite
  storage backends.
\newblock LMSYS Blog, 2025.
\newblock URL \url{https://lmsys.org/blog/2025-09-10-sglang-hicache/}.
\newblock Accessed 2026-05-04.

\bibitem[Xie et~al.(2026{\natexlab{a}})Xie, Huang, and
  Huang]{hisparse_blog_2026}
Zhiqiang Xie, Zhangheng Huang, and Tingwei Huang.
\newblock {HiSparse}: Turbocharging sparse attention with hierarchical memory.
\newblock LMSYS Blog, April 2026{\natexlab{a}}.
\newblock URL \url{https://www.lmsys.org/blog/2026-04-10-sglang-hisparse/}.
\newblock Accessed 2026-05-04.

\bibitem[Xie et~al.(2026{\natexlab{b}})Xie, Xu, Zhao, An, Mailthody, Mahlke,
  Garland, and Kozyrakis]{xie2025strata}
Zhiqiang Xie, Ziyi Xu, Mark Zhao, Yuwei An, Vikram~Sharma Mailthody, Scott
  Mahlke, Michael Garland, and Christos Kozyrakis.
\newblock Strata: Hierarchical context caching for long context language model
  serving.
\newblock In \emph{20th USENIX Symposium on Operating Systems Design and
  Implementation (OSDI)}, 2026{\natexlab{b}}.
\newblock URL \url{https://arxiv.org/abs/2508.18572}.

\bibitem[Yang et~al.(2025)Yang, Li, Yang, Zhang, Hui, Zheng, Yu, Gao,
  et~al.]{qwen3_technical_report}
An~Yang, Anfeng Li, Baosong Yang, Beichen Zhang, Binyuan Hui, Bo~Zheng, Bowen
  Yu, Chang Gao, et~al.
\newblock {Qwen3} technical report, 2025.
\newblock URL \url{https://arxiv.org/abs/2505.09388}.

\bibitem[Yuan et~al.(2025)Yuan, Gao, Dai, Luo, Zhao, Zhang, Xie, Wei, Wang,
  Xiao, Wang, Ruan, Zhang, Liang, and Zeng]{yuan2025nsa}
Jingyang Yuan, Huazuo Gao, Damai Dai, Junyu Luo, Liang Zhao, Zhengyan Zhang,
  Zhenda Xie, Yuxing Wei, Lean Wang, Zhiping Xiao, Yuqing Wang, Chong Ruan,
  Ming Zhang, Wenfeng Liang, and Wangding Zeng.
\newblock Native sparse attention: Hardware-aligned and natively trainable
  sparse attention.
\newblock In \emph{Proceedings of the 63rd Annual Meeting of the Association
  for Computational Linguistics (Volume 1: Long Papers)}, pages 23078--23097,
  2025.
\newblock \doi{10.18653/v1/2025.acl-long.1126}.
\newblock URL \url{https://aclanthology.org/2025.acl-long.1126/}.

\bibitem[{Z.ai}(2026)]{glm52_blog_2026}
{Z.ai}.
\newblock {GLM-5.2}: Built for long-horizon tasks.
\newblock Hugging Face Blog, June 2026.
\newblock URL \url{https://huggingface.co/blog/zai-org/glm-52-blog}.
\newblock Accessed 2026-06-18.

\bibitem[Zhang et~al.(2025)Zhang, Ji, Chen, Fu, Miao, Nie, Chen, and
  Cui]{zhang2025pqcache}
Hailin Zhang, Xiaodong Ji, Yilin Chen, Fangcheng Fu, Xupeng Miao, Xiaonan Nie,
  Weipeng Chen, and Bin Cui.
\newblock {PQCache}: Product quantization-based {KVCache} for long context
  {LLM} inference.
\newblock \emph{Proceedings of the ACM on Management of Data}, 3\penalty0
  (3):\penalty0 201:1--201:30, 2025.
\newblock \doi{10.1145/3725338}.

\bibitem[Zhang et~al.(2023)Zhang, Sheng, Zhou, Chen, Zheng, Cai, Song, Tian,
  R{\'e}, Barrett, Wang, and Chen]{h2o2023}
Zhenyu Zhang, Ying Sheng, Tianyi Zhou, Tianlong Chen, Lianmin Zheng, Ruisi Cai,
  Zhao Song, Yuandong Tian, Christopher R{\'e}, Clark Barrett, Zhangyang Wang,
  and Beidi Chen.
\newblock {H2O}: Heavy-hitter oracle for efficient generative inference of
  large language models.
\newblock In \emph{Advances in Neural Information Processing Systems}, 2023.

\bibitem[Zheng et~al.(2024)Zheng, Yin, Xie, Sun, Huang, Yu, Cao, Kozyrakis,
  Stoica, Gonzalez, Barrett, and Sheng]{zheng2024sglang}
Lianmin Zheng, Liangsheng Yin, Zhiqiang Xie, Chuyue Sun, Jeff Huang, Cody~Hao
  Yu, Shiyi Cao, Christos Kozyrakis, Ion Stoica, Joseph~E. Gonzalez, Clark~W.
  Barrett, and Ying Sheng.
\newblock {SGLang}: Efficient execution of structured language model programs.
\newblock In \emph{Advances in Neural Information Processing Systems}, 2024.
\newblock URL \url{https://openreview.net/forum?id=VqkAKQibpq}.

\bibitem[Zhong et~al.(2024)Zhong, Liu, Chen, Hu, Zhu, Liu, Jin, and
  Zhang]{distserve2024}
Yinmin Zhong, Shengyu Liu, Junda Chen, Jianbo Hu, Yibo Zhu, Xuanzhe Liu, Xin
  Jin, and Hao Zhang.
\newblock {DistServe}: Disaggregating prefill and decoding for
  goodput-optimized large language model serving.
\newblock In \emph{18th USENIX Symposium on Operating Systems Design and
  Implementation (OSDI)}, 2024.

\end{thebibliography}

\appendix

\section{Implementation in SGLang}
\label{sec:appendix-impl}

This appendix summarizes what \sys{} adds to SGLang and where it
attaches to the existing engine. The feature is enabled by a single
flag and otherwise leaves the serving stack unchanged; it comprises
roughly $2{,}200$ lines of new Python across six modules plus a CUDA
kernel header, together with integration changes to the scheduler,
model runner, attention backends, and the disaggregation path.

\paragraphhead{New components} The \emph{coordinator}
(\texttt{managers/hisparse\_coordinator.py}, ${\sim}1{,}000$ lines)
owns the request lifecycle of \S\ref{sec:workflow}: it stages prefill
KV to the host pool, allocates and grows the per-request-per-layer GPU
caches, performs the write-through of newly generated KV, and
orchestrates swap-in, including the plan-then-IO prefetch groups of
\S\ref{sec:prefetch}. The \emph{fused kernels}
(\texttt{jit\_kernel/hisparse.py} with \texttt{csrc/hisparse.cuh})
implement \textsc{Resolve} for token-level and compressed KV
layouts---with optional recording of the miss plan---plus the
copy-only kernel that replays a recorded plan for shared layers. The
\emph{memory layer} (\texttt{mem\_cache/allocator/hisparse.py},
\texttt{hisparse\_memory\_pool.py}, \texttt{pool\_host/hisparse.py})
provides paged host-pool allocation, the device-side cache pool, and a
mixin over SGLang's existing host KV cache class (below). A small
\emph{configuration module} (\texttt{arg\_groups/hisparse\_hook.py})
applies backend defaults and validates attention-backend and KV-dtype
compatibility.

\paragraphhead{Engine integration} Beyond the mechanisms of
\S\ref{sec:design}, the engineering work concentrates in three
places. Concurrency: staging, write-through, and prefetch each run on
their own CUDA stream, ordered against the compute stream by events;
the scheduler admits a request only once its staging acknowledgment
arrives, and its retract and pause paths release \sys{} state
alongside ordinary KV state. Graph capture: \textsc{Resolve} and the
prefetch fork are captured inside SGLang's steady-state decode CUDA
graph, which requires all metadata updates and IO issue to be
replayable without host-side branching. Compatibility: the model
runner auto-enables exact prefetch for models that declare
selection-sharing groups, falls back to synchronous swap-in under
pipeline parallelism or speculative decoding, and a HIP kernel
variant supports AMD GPUs; in disaggregated mode, prefill instances
write KV directly into the decode host's DRAM pool over the existing
transfer backend.

\paragraphhead{Relationship to HiCache} SGLang's
HiCache~\citep{hicache_blog_2025} is a hierarchical \emph{prefix}
cache: host and storage tiers that let finished requests' KV be reused
by later requests with shared prefixes. \sys{} reuses HiCache's
host-tier infrastructure---the pinned host KV pool and its IO
backends---through a mixin, but manages a different object: a
per-request decode working set governed by the model's own sparse
selections. In deployment the two compose as a prefill--decode dual:
HiCache on the prefill node reuses prefixes to cut prefill work, while
\sys{} on the decode node bounds per-request residency to scale decode
batches. The GPU-assisted IO path follows Strata~\citep{xie2025strata}.

\paragraphhead{Configuration} \sys{} is enabled with
\texttt{--enable-hisparse}; a JSON \texttt{--hisparse-config} sets
\texttt{top\_k}, \texttt{device\_buffer\_size} ($B$),
\texttt{host\_to\_device\_ratio} (host-pool capacity relative to the
device KV budget), and the swap-in transfer block size of
\S\ref{sec:kernel}. Exact prefetch enables itself for shared-index
models and can be disabled for ablation via an environment variable;
all experiments in \S\ref{sec:eval} use these switches and no code
changes.

\end{document}